\documentclass[fleqn,usenatbib]{mnras}

\usepackage{newtxtext,newtxmath}

\usepackage[T1]{fontenc}

\DeclareRobustCommand{\VAN}[3]{#2}
\let\VANthebibliography\thebibliography
\def\thebibliography{\DeclareRobustCommand{\VAN}[3]{##3}\VANthebibliography}

\usepackage{graphicx}	
\usepackage{amsmath}	
\usepackage{pdflscape}

\title[Sparse regression modelling of the stellar mass]{A sparse regression approach for populating dark matter halos and subhalos with galaxies}

\author[M. Icaza-Lizaola et al.]{M. Icaza-Lizaola$^{1,2,3}$\thanks{E-mail: miguel.a.de-icaza-lizaola@durham.ac.uk},
Richard G. Bower $^{1,2,4}$,
Peder Norberg$^{1,2,4}$,
Shaun Cole$^{1}$,
\newauthor
Matthieu Schaller$^{5,6}$
\\
\scriptsize $^{1}$Institute for Computational Cosmology, Department of Physics, Durham University, South Road, Durham DH1 3LE, UK.\vspace*{-2pt} \\
\scriptsize $^{2}$Institute for Data Science, Department of Physics, Durham University, South Road, Durham DH1 3LE, UK.\vspace*{-2pt} \\
\scriptsize $^{3}$Korea Astronomy and Space Science Institute,  776 Daedeok-daero,
Yuseong-gu, Daejeon 34055, Republic of Korea.\\
\scriptsize $^{4}$Centre for Extragalactic Astronomy, Department of Physics, Durham University, South Road, Durham DH1 3LE, UK.\vspace*{-2pt} \\
\scriptsize $^{5}$Lorentz Institute for Theoretical Physics, Leiden University, PO Box 9506, NL-2300 RA Leiden, The Netherlands\\
\scriptsize$^{6}$Leiden Observatory, Leiden University, PO Box 9513, NL-2300 RA Leiden, The Netherlands\\
}

\date{Accepted XXX. Received YYY; in original form ZZZ}

\pubyear{2022}

\begin{document}
\label{firstpage}
\pagerange{\pageref{firstpage}--\pageref{lastpage}}
\maketitle

\begin{abstract}

 We use sparse regression methods (SRM) to build accurate and explainable models that predict the stellar mass of central and satellite galaxies as a function of properties of their host dark matter halos. SRM are machine learning algorithms that provide a framework for modelling the governing equations of a system from data. In contrast with other machine learning algorithms, the solutions of SRM methods are simple and depend on a relatively small set of adjustable parameters.
We collect data from 35,459 galaxies from the EAGLE simulation using 19 redshift slices between $z=0$ and $z=4$ to parameterize the mass evolution of the host halos. Using an appropriate formulation of input parameters, our methodology can model satellite and central halos using a single predictive model that achieves the same accuracy as when predicted separately. This allows us to remove the somewhat arbitrary distinction between those two galaxy types and model them based only on their halo growth history. Our models can accurately reproduce the total galaxy stellar mass function and the stellar mass-dependent galaxy correlation functions ($\xi(r)$) of EAGLE. We show that our SRM model predictions of $\xi(r)$ is competitive with those from sub-halo abundance matching and  might be comparable to results from extremely randomized trees. We suggest SRM as an encouraging approach for populating the halos of dark matter only simulations with galaxies and for generating mock catalogues that can be used to explore galaxy evolution or analyse forthcoming large-scale structure surveys.
\end{abstract}




\section{Introduction}

Within the $\Lambda$-CDM paradigm \cite[e.g.][]{planck2013}, an expanding universe filled with particles that interact only through gravity can be accurately modelled using N-body simulations \cite[e.g.][]{2005Natur.435..629S}. Because of advances in computational methods, such simulations can track the formation of galaxy-scale dark matter haloes within volumes approaching the size of the observable Universe. However, these simulations do not include the baryonic component that leads to the formation of stars and galaxies. Hydrodynamical simulations that include baryons need to deal with complicated cooling and feedback processes and are strongly influenced by events happening at scales much smaller than the size of the simulation grid. This makes them significantly more expensive to run and limits their volume to about 1 $\rm Gpc^3$ \citep[e.g.][]{Springel_2017}. There is, therefore, an incentive for a hybrid approach, in which one uses hydrodynamic simulations to learn the relation between dark matter and baryonic tracers, and then uses these relations to populate N-body mock catalogues of larger volume.  
 
In \cite{icazalizaola2021sparse} we present a novel methodology that uses Sparse Regression Methods \citep[SRM;][]{doi:10.1111/j.2517-6161.1996.tb02080.x,10.5555/2834535}  to model the relations between the stellar mass of a galaxy and its host halo in the Evolution and Assembly of Galaxies and their Environments \citep[EAGLE,][]{2015MNRAS.446..521S,2015MNRAS.450.1937C,McAlpine_2016} 100~Mpc hydrodynamical simulation. SRM are a set of machine learning algorithms designed to identify the parameters that better describe a dependent variable, then discard the remaining unnecessary ones. Recently they have been suggested as the appropriate framework to extract the equation of states of a physical system from collected data and with minimal knowledge of the physics of the system \citep{Brunton3932}.

In \cite{icazalizaola2021sparse} we were interested in developing and testing the methodology in a simple scenario without going into some of the more complicated challenges that populating a realistic N-body mock accurately would require. With that in mind, we tested our methodology on central galaxies (the main galaxy within each dark matter halo) only as they have monotonic growths with time which makes them easier to model. In this work, we extend our methodology to include satellite galaxies as well. Satellite galaxies (and their associated dark matter subhaloes) are created when a smaller dark matter halo is accreted by a larger one. This is a common process in the $\Lambda$-CDM model.
As they orbit within the larger halo, satellite galaxies (and their remnant dark matter subhaloes) undergo a much more diverse range of physical processes than their central galaxy counterparts.  

Unlike the main dark matter halo, which undergoes monotonic mass growth, the remnants of smaller accreted haloes may decay with time \citep[e.g.][]{2004cgpc.symp..325B,2018MNRAS.474.3043V} as they loose mass due to processes such as tidal stripping and heating \citep{1967MNRAS.136..101L,1983ApJ...264...24M,2003ApJ...584..541H,Green_2019}. Moreover, the satellite galaxies residing inside these remnant halos are subject to `environmental' processes that remove cold gas and suppress the accretion of more material \citep{1972ApJ...176....1G,Vollmer_2001, larson1980, bahe2015, correa2019}. As a result, star formation in satellite galaxies is significantly suppressed compared to central galaxies and we expect less stellar mass growth.

In EAGLE, the differentiation between central halos and subhalos is done by the SUBFIND algorithm \citep{2001MNRAS.328..726S}. Within each halo, the algorithm identifies the self-bound overdensities and classifies them as independent subhalos. The subhalo with the lowest potential energy is classified as the central halo and assigned any diffuse mass that has not already been associated with a subhalo. This distinction is made separately at each output time and is not a fundamental differentiation, but dependent on the details of the algorithm. In some cases, this leads to anomalous behaviour, in particular inconsistent classifications of the same subhalo at different redshift slices \citep[e.g.][]{2015MNRAS.454.3020B}. It is, therefore, desirable to use a methodology that does not make a fundamental distinction between central and satellite galaxies when modelling the stellar mass, but rather to use the same approach based on the overall halo mass history.

In this paper, we use a lower threshold in the host halo mass for our central galaxy sample compared to \cite{icazalizaola2021sparse}, reducing it from $M=10^{11.1}{\rm M_\odot}$ to $M=10^{10.6}{\rm M_\odot}$. This allows us to identify low mass haloes which contain relatively large galaxies (with stellar masses greater than $10^{9}{\rm M_\odot}$). This is a particularly important consideration for satellite galaxies, if we are to generate a stellar-mass complete catalogue. 

Other works have used machine learning algorithms to model the relationship between the halo and stellar properties inside a hydrodynamical simulation \citep[e.g.][]{10.1093/mnras/stv2981,10.1093/mnras/sty1169}. Their models accurately reproduce several statistics of the original simulation. However, given that these types of models generate {\it black box} answers it might be complicated to modify them to reproduce statistics from observations instead. \cite{lovell2021machine} trains an extremely randomized tree \citep{ERT_paper} model on data from the EAGLE simulation and uses it to populate the P-Millennium N-body simulation with galaxies \citep{10.1093/mnras/sty3427}. \cite{Moster_2021} uses a neural network approach that rewards the algorithm for reproducing observed statistics of a survey (like correlation functions and stellar mass functions) instead of properties of individual galaxies. This circumvents the problem of differences in statistics between the hydrodynamical simulation used to calibrate the model and those from an observational survey, at the cost of not requiring accuracy in the predictions of the individual values of galaxy properties. Given that our model is an equation of state with a set of input parameters fitted by the model, it is in principle possible to extract the best advantages of both approaches, extracting the important physical parameters by comparison to the simulation, but optimising the coefficients of these terms to reproduce the statistics of an observational data set.

This paper is organized as follows. Section~\ref{methodology} summarises the sparse regression methodology used in this work, with a complete discussion of the methodology presented in \cite{icazalizaola2021sparse}. Section~\ref{sec:data} introduces the data set that we use and any enhancements to the model that we have made to handle the more complex data-set. In particular, \S\ref{subsec:matching} explains the details of the bijective match between the hydrodynamical EAGLE simulation and the EAGLE dark matter only (EAGLE DM hereafter)  simulation. \S\ref{subsec:mass_methodology} and \S\ref{weights_section} describe the methodology used to extract our training data set from the EAGLE DM only simulation as well as the new parametrisation of the model and the new weighting scheme adopted. The results from our different models are shown and analysed in Section~\ref{sec:results}. In Section \ref{Cluster_&_SM} we compare the stellar mass function and the clustering of our resulting models with the ones from the original EAGLE sample. In Section \ref{other methods} we compare our resulting models with some available from the literature. Our conclusions and thoughts on the potential of the current methodology are discussed in Section~\ref{conclusions}.

\section{Methodology}
\label{methodology}

The methodology followed in this work is presented in detail in~\cite{icazalizaola2021sparse}. Here we include a summary of the key concepts, and then in Sections~\ref{subsec:mass_methodology} and~\ref{weights_section}, we describe the additions and changes to the methods adopted in this specific work.

Sparse Regression Methods \citep[SRM;][]{doi:10.1111/j.2517-6161.1996.tb02080.x,10.5555/2834535,2017arXiv171200484T} are a set of machine learning algorithms designed to develop a fitting function by selecting linear combinations from a large library of candidate functional forms. The method selects only a minimal subset of functions from the library such that the combination describes the input data well but does not over-fit and hence avoids poor interpolation between input points. One key advantage of SRM methods over other machine learning techniques is that the resulting model is in the form of an equation with nominally a small subset of terms, making it more likely to have a clear physical interpretation.

 In \cite{icazalizaola2021sparse} we used SRM to model the relation between the stellar mass ($M^*$) of central galaxies and a set of properties of their host halos. We found that a good, but simple description could be obtained based on the final mass of the host halo and its parameterized formation history. In this paper, we aim to provide a similar relationship that describes all galaxies in the simulation, whether they are the dominant galaxy within the halo (which we refer to as {\it central}) or a galaxy that was formed in a separate sub-halo that has been subsequently been accreted (we refer to such galaxy as a {\it satellite}). Although we follow very similar methodologies to \cite{icazalizaola2021sparse}, the halo and sub-halo properties used as input parameters here have been adapted so that we can model both satellites and central galaxies consistently. The details are described in Section~\ref{subsec:mass_methodology}.
 
 Let us call $M$ the number of host halo properties, and $N$ the total number of galaxies in our data set. For each halo property we define the vector $\vec{x}'_i=[x'_{1i}, ... ,x'_{Ni}]$ that contains the observed value of the $i^{\rm th}$  property of each host halo ($i\leq M$). 
 
The input halo properties need to be standardised so that they all vary within a consistent range. This is done using the following transformation
 \begin{equation}
 \label{normalisation}
 \vec{x_i}=\frac{\vec{x'_i}-\mu(\vec{x'_i})}{\sigma({\vec{x'_i}})}
 \end{equation}
where $\mu$ and $\sigma$ are the mean and standard deviation operators respectively.   
We use the  $\vec{x}_i$ vectors to build a set of $D$ polynomial functions $F_l(\vec{x})$ ($l<D$), where the function $F_l$ can be either a linear, a quadratic or cubic combination of the dependent variables, i.e.\  $F_{l\alpha}=x_{i\alpha}$ or $F_{l\alpha}=x_{i\alpha} \times x_{j\alpha}$ or $F_{l\alpha}=x_{i\alpha} \times x_{j\alpha} \times x_{k\alpha}$, where $1 \le i \le j \le k \le M$ and $\alpha<N$. We use all possible linear quadratic and cubic combinations of the input properties and so $D= 1 +M + M(M+1)/2 +M(M+1)(M+2)/6$.
 
The value of the stellar mass predicted by our model for galaxy $\alpha$ ($M^*_{p\alpha}$)\footnote{
As mentioned later, our code actually models $\log_{10}(M^*/M_\odot)$. We opt against including the full logarithmic expression in the main text of the paper and associated equations to simplify the notation, while we show the explicit dependencies in the figure labels.
}
is expressed as the linear combination of functions $F_{l\alpha}$: 
 \begin{equation}
 M^*_{p\alpha}= \sum_{l=0}^{D} C_l F_{l\alpha}
 \end{equation}
where $\vec{C}=[C_0,..,C_D]$ are a set of coefficients. The optimal values of these coefficients are the quantities determined by our methodology. 
 
Following the SRM approach, most coefficients are discarded (i.e., we set $C_l=0$) and only a small subset of the possible coefficients are retained. This is achieved by minimising the LASSO function defined as:
\begin{equation}
\label{LASSO}     
 L(\vec{C})=\chi^2(\vec{C})+\lambda P(\vec{C})
\end{equation}
where $\chi^2(\vec{C})$ is a statistic that determines the goodness of the fit, $P(\vec{C})$ is a penalty term that incentivises the minimisation to discard unnecessary input parameters and $\lambda$ is a hyperparameter of our methodology that regulates the relative magnitude of $P(\vec{C})$. We define $\chi^2(\vec{C})$ as:
\begin{equation}
\label{chi^2} 
\chi^2(\vec{C})= \sum_{\alpha=1}^N\frac {(M_\alpha^*-M_{p\alpha}^*(C))^2}{\sigma^2},
\end{equation}
where $\sigma$ is an estimate of the uncertainty of the measurement of $M_\alpha^*$ (as defined by equation~11 of \cite{icazalizaola2021sparse}). 

The penalty term $P(\vec{C})$ is defined in such a way that its value increases significantly with the number of coefficients $C_j$ that are non-zero. The shape of $P(\vec{C})$is given by the following equation:
\begin{equation}
\label{Penalty}
P(\vec{C})=\sum_{l=1}^D  \left[\sum_{m\neq l} \mid C_m \mid e^{-\left( {\epsilon}/{C_m} \right)^2} \right] \mid C_l \mid e^{-\left( {\epsilon}/{C_l} \right)^2},
\end{equation}
where $\epsilon$ is a small constant that determines how close to zero a coefficient needs to be before its contribution to the penalty is negligible. 

In this work, all coefficients below $10^{-3}$ are discarded. We refer the reader to \cite{icazalizaola2021sparse} for a discussion of the choice of this specific value and of the optimal $\epsilon$ value for third order polynomials.

Equation~\ref{LASSO} is designed to avoid overfitting the input data, which is a necessity in any model with a large space of input parameters. This is achieved by the balancing between the goodness of fit and the penalty term. An overfitted model would have a small $\chi^2$ by including many non-zero parameters, which, in turn, would make the penalty term very large. Therefore the minimum of $L(\vec{C})$ should correspond to a model that is as simple as possible (small $P(\vec{C})$) while still being a good fit (small $\chi^2$). The equilibrium between the need to fit the data well and to keep the number of non-zero coefficients small is set by the choice of the $\lambda$ parameter: a large value strongly reduces the number of coefficients selected, while a small value does not penalize the goodness of fit enough. We determine the optimal value using the k-fold methodology \citep{10.5555/2834535}, where the data is separated into a training set and a test set k-times. The optimal value of $\lambda$, and its associated uncertainty, can then be determined by examining how well a model fitted to the training set can predict the data in the test set. The full details of this process are described in \cite{icazalizaola2021sparse}.

We use 85$\%$ of our data to train our model, with the remaining 15$\%$ labeled as the Holdout data set. The latter is used in section~\ref{results_main} to test the accuracy of the method, while the full data set is used in sections~\ref{Cluster_&_SM} and beyond.

\section{Data}
\label{sec:data}

Our data set comes from the EAGLE \citep{2015MNRAS.446..521S,2015MNRAS.450.1937C,McAlpine_2016} simulations, which are a suite of hydrodynamical simulations built using the Planck 2014 cosmology \citep{planck2013}. During the rest of this work we define the stellar mass of a galaxy in EAGLE as the sum of all stellar particles inside a sphere with an aperture of 30 kpc centered at the center of the potential of the galaxy.

We use the simulations built in a 100 comoving Mpc box, which is the largest box available. Halos in the simulation are identified using a  Friends-of-Friends algorithm \citep[FoF; e.g.][]{1985ApJ...292..371D} with a linking length of $b$=0.2. Subsequently, the SUBFIND algorithm \citep{2001MNRAS.328..726S} finds the subhalos within each halo and selects one of them as the central halo. The simulation outputs are saved in 29 snapshots going from $z$=20 to $z$=0. The snapshots are used to build merger trees \citep{2017MNRAS.464.1659Q} by identifying halos with their progenitors at the previous redshift slice. Main progenitors are defined as the progenitors with the larger branch mass \citep{De_Lucia_2007}, defined as the sum of the progenitors mass at all previous snapshots. During this work, we use the main progenitor branch to track the mass evolution of a halo.

\subsection{Matching}
\label{subsec:matching}
The goal of this work is to develop a fitting function that allows the mass of a galaxy to be estimated from knowledge of its DM halo formation history only.
Since DM halos in hydrodynamical simulations are affected by baryonic processes that might alter their density profile \citep{10.1093/mnras/stv1341,10.1111/j.1365-2966.2012.20879.x,1996MNRAS.283L..72N}, or other properties like the shape of the halo \citep{1991ApJ...377..365K,2013MNRAS.429.3316B}, it is important that we match the haloes in the hydrodynamical simulation with the same haloes in a dark matter only simulation (with identical cosmology and initial conditions). By making a one-to-one matching between the DM only simulation and the hydrodynamical one, the properties of the DM only simulations can be used as the input variables of the model (the vectors $\vec{x'}_j$ of Section~\ref{methodology}) while the stellar mass is measured in the full-physics hydrodynamical simulation.
The matching is done by following the procedure of \cite{10.1093/mnras/stv1067}. To summarise, we look at the 50 most bound DM particles of each halo or subhalo in the hydrodynamical simulation: if a halo or subhalo of the DM only simulation contains at least half of these particles, then they are matched. The matching is done for all halos above $M_\text{{total}}> 2 \times 10^9 {\rm M_\odot}$ and both halos need to be above this value to be matched, where $M_\text{{total}}$ is the summed mass of all particles assigned to the halo or subhalo.

\begin{figure}
\includegraphics[width=85mm,height=85mm]{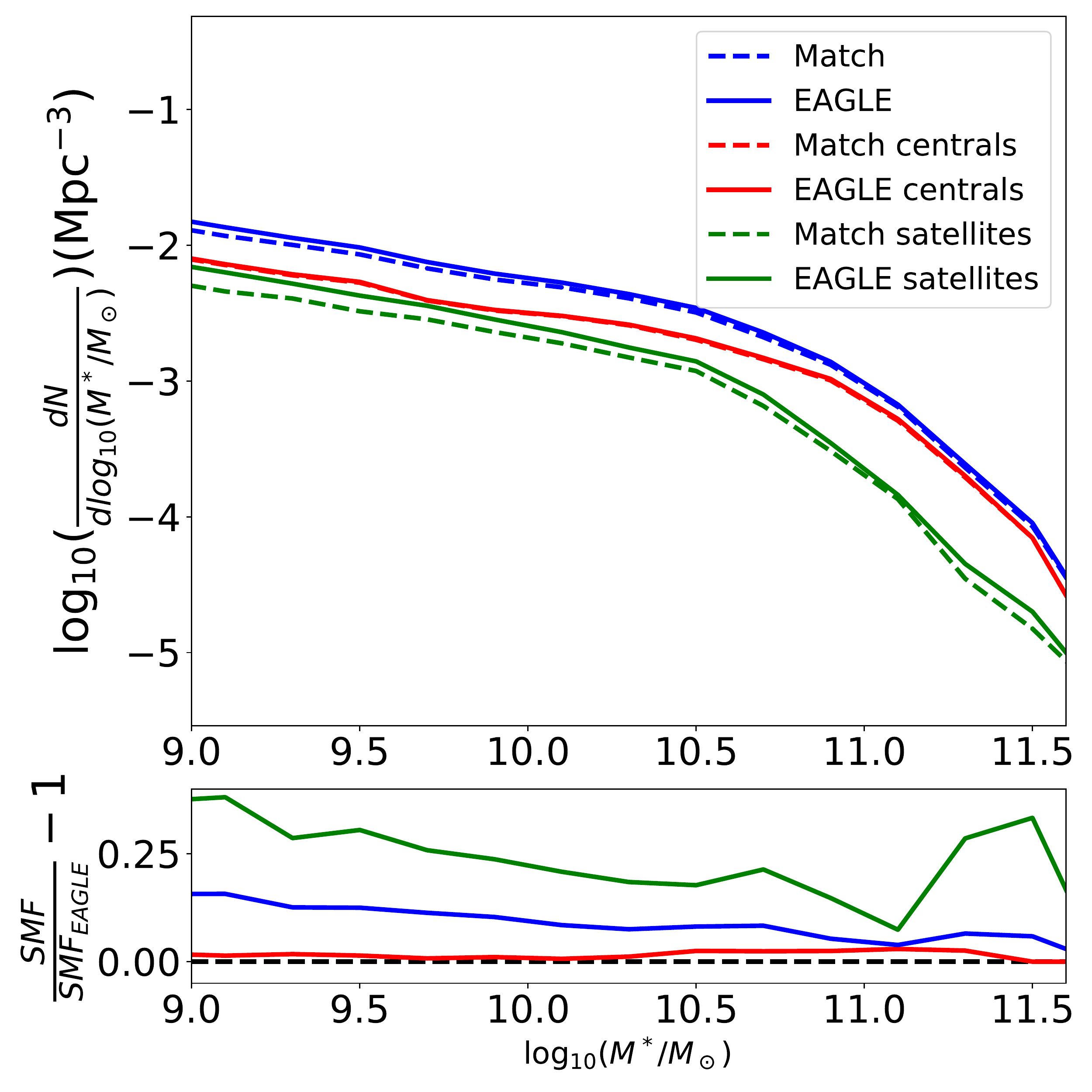}
\caption{Comparison of the Stellar Mass Function (SMF) of the full EAGLE simulation (solid lines), with the SMF from the galaxies living in halos that were successfully matched (dashed lines). The plot shows results for both central halos (red) and satellites (green), and the combined sample of central and satellites halos (blue). The bottom panel shows the ratios of the SMF of comparable galaxy types (while keeping the colour coding the same as in the top panel) and quantifies the fraction of matching failures per galaxy type.}
\label{Matching_SMF}
\end{figure}

Fig.~\ref{Matching_SMF} shows the stellar mass function (SMF) of the full EAGLE hydrodynamical simulation and compares it to the SMF of the galaxies living in halos that were successfully matched. The bottom panel of Fig.~\ref{Matching_SMF} shows that the fraction of matching failures for central galaxies is around $1\%$ for all stellar mass scales of interests. This explains why it was not necessary to consider the effect of unmatched haloes in \cite{icazalizaola2021sparse}. However, the number of unmatched satellite galaxies is significantly larger, with a matching success rate around $80\%$ for galaxies with $\log_{10}(M^*/{\rm M_\odot})>10$ (green line in the bottom panel of Fig.~\ref{Matching_SMF}).

With this in mind, all statistics presented from Section~\ref{Cluster_&_SM} onwards result from applying the model to all halos in the EAGLE DM only simulation (matched and unmatched) and compares them to statistics from all galaxies in the hydrodynamical simulation. This comparison assumes that the distribution of unmatched halos in both simulations is similar. We explore the validity of this assumption in Appendix~\ref{Matching_failures}.

\subsection{Halo Selection and Input Parameterisation}
\label{subsec:mass_methodology}

We begin our selection of haloes by tracing the evolution of the halo mass in the DM only simulation at 19 redshift slices between $z=0$ and $z=4$. This initial selection is based on $M_\text{{total}}(z)$, the total mass of the particles associated to the halo or sub-halo by the SUBFIND algorithm. These trajectories summarise the evolution of the galaxies host halo mass as a function of redshift and give us a relation between halo mass and time for each galaxy. In order to ensure that the trajectory is not overly affected by the algorithm used in the selection process, we use a Gaussian kernel with a $\sigma$ of one redshift slice to smooth this evolution history. Since halo masses can increase as well as decrease (for satellite galaxies in particular), we base our halo selection on the maximum value of $M_\text{{total}}(z)$ in the smoothed trajectory. The success rate of the matching is dependent on the halo mass, with more massive halos being more likely to be matched.
We find that $\mathbf{Max}(M_\text{{total}}(z))=10^{10.66}{\rm M_\odot}$, corresponds to the threshold at which more than 90~per~cent 
of halos are successfully matched.
We define this threshold as the halo mass cutoff of our sample. In order to avoid missing data, we discard those that do not have a well-defined main progenitor in all redshift slices up to $z=4$. For $\mathbf{Max}({M_\text{{total}}(z)})>10^{10.66}{\rm M_\odot}$, this cut is unimportant, with 99.6~per~cent of the sample being kept. Our final sample consists of a total of 35,456 galaxies, of which 9,967 live inside subhalos, and 25,489 inside central halos.

As a pre-processing step, we use the interpolation scheme developed in \cite{icazalizaola2021sparse} to ensure the halo masses of central galaxies are not affected by inconsistent classification between snapshots.
Nominally halos in our models have their evolution tracked with $M_\text{{total}}(z)$ at all redshifts. We have compared models with different halo mass definitions for centrals, like $M_{200}^c$\footnote{The mass within a radius for which the density is 200 times larger than the critical density of the Universe. We note that $M_{200}^c$ is only defined for central galaxies in EAGLE.}, and found negligible differences on the accuracy of the stellar mass predictions.

\begin{figure}
\centering
\includegraphics[width=85mm,height=75mm,trim={0 0 0 0.45cm},clip]{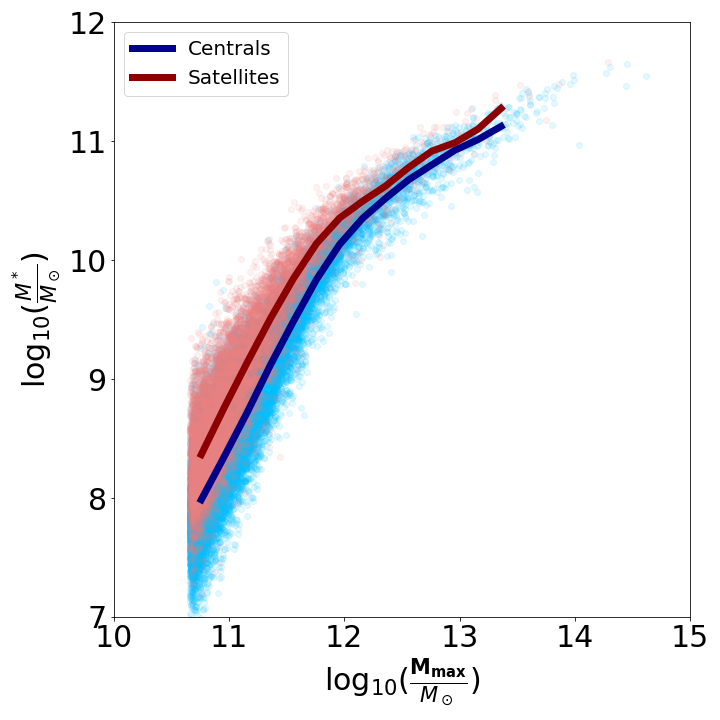}
\caption[Distribution of centrals and satellites in the $M_{\rm max}$-$M^*$ space]{Distribution of the central galaxies (blue dots) and satellite galaxies (red dots) in our sample in the $M_{\rm max}$-$M^*$ space, where $M_{\rm max}$ is the largest halo mass the halo's main progenitor reached (see Eq.~\ref{eq:m_max}). The solid lines show the median value of the distributions. The plot shows that at a fixed  $M_{\rm max}$ the median galaxy mass of a satellite galaxy is larger than that of a central galaxy.
}
\label{Data_cuts}
\end{figure}

Since the satellite halo mass cannot be expected to grow monotonically with decreasing redshift, a more important parameter for each galaxy is instead its maximum halo mass. In the rest of the paper, we refer to this as $M_{\rm max}$:

\begin{equation}
M_{\rm max} = \mathbf{Max}(M_\text{{total}}(z))
\label{eq:m_max}
\end{equation}

Central galaxies tend to grow monotonically with time, and $M_{\rm max}$ is correlated with the stellar mass through the $z=0$ stellar mass - halo mass (SMHM) relation. In satellite galaxies, however, $M_{\rm max}$ corresponds to the redshift at which their host halo merges and becomes the subhalo of a larger system. Once a halo merges the mass of the halo declines due to tidal processes. We can expect, therefore, that the galaxy mass at $z=0$ will be well correlated with the mass of the host halo before merging. 
Fig.~\ref{Data_cuts} shows the distribution of galaxies in the $M_{\rm max}$-$M^*$ space. 

We note that the median stellar mass of satellite galaxies is larger than that of centrals at fixed $M_{\rm max}$, i.e.\ for a fixed $M_{\rm max}$ satellite galaxies are more massive. The offset in the SMHM relation for satellites and centrals is driven by two competing processes. On the one hand, satellites may undergo a strong suppression of their star formation as they orbit within the main halo due to the combined effects of ram-pressure stripping (the removal of the interstellar medium of the galaxy by ram pressure) and {\it strangulation} (the absence of gas infall onto the satellite). On the other hand, while the halo mass of the central continues to grow with cosmic time, the satellite reaches its peak mass and $M_{\rm max}$ becomes frozen thereafter. The net offset is determined by whether the halo mass or the stellar mass grow fastest in the central galaxies, and by whether satellite galaxies are able to continue to grow in stellar mass after they are accreted \citep{behroozi2019}. Because the effect on the stellar mass growth tends to be delayed compared to the effect on the halo, satellite galaxies tend to have larger stellar mass than their central counterparts.

We now describe the input parameters used in this work, which are the values of the vectors $\vec{x'}_j$ of Section~\ref{methodology}. 

In~\cite{icazalizaola2021sparse}, we tested different parameterisations and concluded that parameters that measure the SMHM relation and the halo growth trajectory are the most useful for modelling the stellar mass at $z=0$. We also found no improvement in our models when adding parameters correlated with the angular momentum evolution of the halo. The best model that we found used $\log_{10}(M_{200}^c(z=0)/M_{\odot})$ as the input parameter that traced the SMHM relation, as well as a set of formation criteria parameters $\mathbf{FC}_p$ that model the assembly history, where $\mathbf{FC}_p$ is the redshift by which a central galaxy has assembled $p=[20,30,50,70,90]$~per~cent of its current mass. In order to accommodate satellite galaxies, we substitute the input parameter $M_{200}^c(z=0)$ with $M_{\rm max}$ and we define the dimensionless parameter
\begin{equation}
    {\rm lgM_{ max}}=\log_{10}(M_{\rm max}/{\rm M_\odot})
\end{equation}
and redefine the formation criteria parameters $\mathbf{FC}_p$ as follows. First we find the redshift $z_i$ at which a halo or subhalo reaches $M_{\rm max}$. Then we look at the evolutionary history of the halo from $z=4$ up until $z_i$, and find the redshift ($z_i \le \mathbf{FC}_p \le z=4$) at which the halo has assembled a percentage $p$ of $M_{\rm max}$.

Note that if $z$ is such that $M(z)=M_{\rm max}$, then $z<\mathbf{FC}_{90}<\mathbf{FC}_{70}$.
This parameterisation is almost equivalent to the one used in \cite{icazalizaola2021sparse} when only considering central galaxies as in this case $M_{200}^c(z=0)\sim M_{\rm max}$. As a check, we ran our methodology on the data set of \cite{icazalizaola2021sparse} with the new parameterisation. The resulting model is comparable to the original one in accuracy and simplicity.
In total we use six independent variables in our methodology [${\rm lgM_{ max}}$, $\mathbf{FC}_{20}$, $\mathbf{FC}_{30}$, $\mathbf{FC}_{50}$, $\mathbf{FC}_{70}$, $\mathbf{FC}_{90}$]. 
Each of these parameters is transformed to the standardised space defined by equation~\ref{normalisation}. 
Since we consider cubic combinations of these parameters this leads to a model with up to $D=84$ parameters.

Many methodologies have found that parameters related to the circular velocity of halos, like the maximum of the
radial circular velocity profile at $z=0$ ($V_\text{max}$) or even the maximum value of $V_\text{max}$ among all redshifts ($V_{\rm peak}$), are more accurate than the halo mass when modeling the stellar mass of their host galaxy \citep[e.g.][]{2006ApJ...647..201C,2016MNRAS.460.3100C,2017MNRAS.465.2381M,10.1093/mnras/stv2981,lovell2021machine}. In our current implementation, strongly correlated parameters that serve a similar function in the modelling of the stellar mass, like $V_\text{max}$, $V_{\rm peak}$ and $M_\text{total}$, are not easily distinguished by our algorithm. This leads to subtle variations in the
surviving parameters of a given model that can depend on configuration parameters, like the starting
point of the minimization and the specific training set selection. Degeneracies due to correlated model parameters are further discussed in  \cite{icazalizaola2021sparse} and at the end of Section~\ref{sec:results}.

We have run a model where we use both $M_{\rm max}$ and $V_{\rm peak}$ as free parameters simultaneously, and compare it to the model with only $M_{\rm max}$ that we present in the next section. We found no difference in accuracy or simplicity between the two models. However, the fact that one model is a function of both parameters made its interpretation less straightforward. For example, when running our algorithm using only $M_{\rm max}$, the SMHM relation is modelled as a third-order polynomial of $M_{\rm max}$ (as we show in Section \ref{results_main}), which makes intuitive sense when looking at Fig.~\ref{Data_cuts}. However, when using both  $M_{\rm max}$ and $V_{\rm peak}$, the SMHM function is now modelled by a more complicated function of both parameters. Therefore, by adding parameters that are strongly correlated with $M_{\rm max}$, we lose explainability without gaining accuracy, and hence we decide to keep only one of the two correlated parameters. In Appendix~\ref{Vpeak} we discuss why we did select $M_{\rm max}$ instead of $V_{\rm peak}$. A possibility to work with correlated parameters without the need of doing this sort of correlation analysis beforehand would be to use some principal component analysis  \citep[e.g.][]{doi:https://doi.org/10.1002/0470013192.bsa501}.

To test the differences between modelling satellite and central galaxies separately and modelling them together with a single model, we run three models independently of each other:
\begin{itemize}
    \item A model that only contains central galaxies, with $N=25,489$ data points.
    \item A model that only contains satellite galaxies, with $N=9,967$ data points.
    \item A model that combines central and satellite galaxies and fits them all at the same time, with $N=35,456$ data points.
\end{itemize}

\subsection{Weighting the Cost Function}
\label{weights_section}

In~\cite{icazalizaola2021sparse}, we used a simple $\chi^2$ measure to assess the quality of the model's prediction of the data (i.e.\ $\chi^2$ is the cost function).
In the CDM paradigm, however, smaller halos are always much more numerous than massive ones. As a consequence, such methodology would have a stronger incentive to fit numerous smaller halos more accurately at the expense of a less accurate fit to less numerous massive ones. In~\cite{icazalizaola2021sparse}, we concluded that our methodology became more inaccurate for galaxies larger than $\log_{10} (M^*/{\rm M_\odot})>11.0$ (see discussion of Fig.~14) due to a relatively small fraction of galaxies above the threshold (90 out of $\sim$9,500). Given that in this iteration of the work we reduced the cutoff value of galaxies even further, we now have a larger number of smaller galaxies making the issue even more problematic.
A good solution to this problem is to assign a weight $w'_i$ to each halo. This weight determines how much of an incentive the code will have to fit a particular halo mass correctly. If the weight $w'_i$ is larger for galaxies in larger halos, then by modifying Eq.~\ref{chi^2} to include a normalised weight $w_i$ as below, we will give a larger importance to the rarer larger haloes:
\begin{equation}
\label{chi^2_w} 
\chi_w^2= \sum_{\alpha=1}^N\frac {w_\alpha(M_\alpha^*-M_{p\alpha}^*(C))^2}{N^2}
\end{equation}

To compute the weight of a halo we first look at the halo mass function (HMF) as a function of ${\rm lgM_{ max}}$. To avoid noisy weights from having a small number of objects in the more massive bins, we make use of a linear fit to the HMFs. Referring to the linear fits as $\textrm{fl}({\rm lgM_{ max}})$, the weight of a halo is defined as:
\begin{equation}
w'_\alpha=\sqrt{\frac{10^{\textrm{fl}(\mu)}}{10^{\textrm{fl}({{\rm lgM_{ max}}}_\alpha})}}
\end{equation}
where $\mu$ is the median value of ${\rm lgM_{ max}}$. As a final step, we normalize the weights of a sample as follows
\begin{equation}
\label{weights}
w_\alpha=\frac{N \times w'_\alpha}{\sum_{\alpha=1}^N (w'_\alpha)} 
\end{equation}

We emphasise that in the combined model, the weighting scheme does not distinguish between central and satellite galaxies.

\section{Results}
\label{sec:results}

We start in Section~\ref{results_main} by comparing input and predicted stellar masses, using the holdout data only. 
As mentioned in Section~\ref{methodology}, halos in the holdout set were not used to train the model. Therefore comparisons with the holdout data enables the accuracy of our method to be tested by making model predictions on EAGLE data that the model has not seen before.
In section~\ref{Cluster_&_SM} we present model predictions using the full data set for the galaxy stellar mass function and galaxy clustering split by stellar mass. In Section~\ref{other methods} we compare our EAGLE SRM predictions with a SHAM model \citep{2016MNRAS.460.3100C} and a ML method \citep{lovell2021machine} applied to EAGLE as well.
In section~\ref{OtherParameters}, we consider whether some of the additional parameters identified by the two aforementioned papers could improve our model.

\subsection{Comparing input and predicted stellar masses}
\label{results_main}

\begin{table}
\begin{tabular}{|c|c||c||c|}
 \hline

Coefficient & Centrals & Satellites & Combined\\
\hline

\hline
Constant & 0.122 & 0.172 & 0.171\\ 
${\rm lgM_{ max}}$ &  1.20 & 1.12  & 1.17\\ 
${\rm (lgM_{ max}})^2$& -0.144 & -0.154  & -0.146\\ 
${\rm (lgM_{ max}})^3$ & 0.00527  & 0.00633 & 0.00509\\
$\mathbf{FC}_{20}$ & 0.0435 &  -  & 0.0136\\ 
$\mathbf{FC}_{30}$ & - &  -  & 0.0223\\ 
$\mathbf{FC}_{50}$ & -0.0732  &  0.0603  & 0.0560\\ 
$\mathbf{FC}_{70}$ & 0.0803 &  0.110  & 0.0953\\ 
$\mathbf{FC}_{90}$ &  0.0262 & 0.100 & 0.190\\
$(\mathbf{FC}_{30})^2 $ & - & - &0.0107\\ 
${\rm lgM_{ max}} \times \mathbf{FC}_{20}$ & -0.0392 &-  &-0.0224\\
${\rm lgM_{ max}} \times \mathbf{FC}_{30}$ & - & - &-0.00508\\
${\rm lgM_{ max}} \times \mathbf{FC}_{50}$ & - & -0.0595 &-0.0263\\
$\mathbf{FC}_{20}\times \mathbf{FC}_{90}$ & - & - & -0.0220 \\ 
$\mathbf{FC}_{30}\times \mathbf{FC}_{90}$ & - & - & -0.0192 \\
$\mathbf{FC}_{50}\times \mathbf{FC}_{90}$ & - & -0.0450 & -0.0636 \\
$\mathbf{FC}_{70}\times \mathbf{FC}_{90}$ & - & -0.0121 & - \\
$(\mathbf{FC}_{20})^3$ & 0.0106 & - & - \\
$(\mathbf{FC}_{30})^3$ & 0.00521 & - & - \\
${\rm (lgM_{ max})}^2 \times \mathbf{FC}_{30}$ & - & -0.00217 & - \\
${\rm (lgM_{ max})}^2 \times \mathbf{FC}_{90}$ & - & -0.00521  & -0.0124 \\
${\rm lgM_{ max}} \times (\mathbf{FC}_{20})^2$ & - & 0.00197 & - \\
${\rm lgM_{ max}} \times (\mathbf{FC}_{90})^2$ & - & - & -0.00433 \\
$(\mathbf{FC}_{20})^2\times \mathbf{FC}_{70}$ &  -& 0.00567 & 0.00875\\ 
$(\mathbf{FC}_{30})^2 \times \mathbf{FC}_{20}$ & - &-  & -0.00186 \\ 
$(\mathbf{FC}_{50})^2\times \mathbf{FC}_{20}$ & -0.00158 & - & 0.0243 \\ 
\end{tabular}
\caption[Parameters and their respective coefficients]{Parameters and their respective values for the surviving coefficients of the three models. Note that the parameters presented here are in the standardised space defined by Eq.~\ref{normalisation}. Parameters are shown to three significant figures, sufficient to make the RMSE accurate to four significant figures.}
\label{Results_table}
\end{table}

We now present the results of each of our three models. 
The surviving coefficients and their respective values are shown in Table~\ref{Results_table}. 
In order to extract a fitting function that can be applied directly to the input variables, one first needs to transform the input data using Eq.~\ref{normalisation}, which requires the mean and standard deviation values of the dependent variables.  The values of these parameters for our combined model are given in Table~\ref{Normalisation_table}\footnote{Note that the resulting stellar mass also needs to be converted from standardised units, and we have therefore included the stellar mass parameters in Table~\ref{Normalisation_table} as well.}.

\begin{figure*}
\includegraphics[width=85mm,height=85mm]{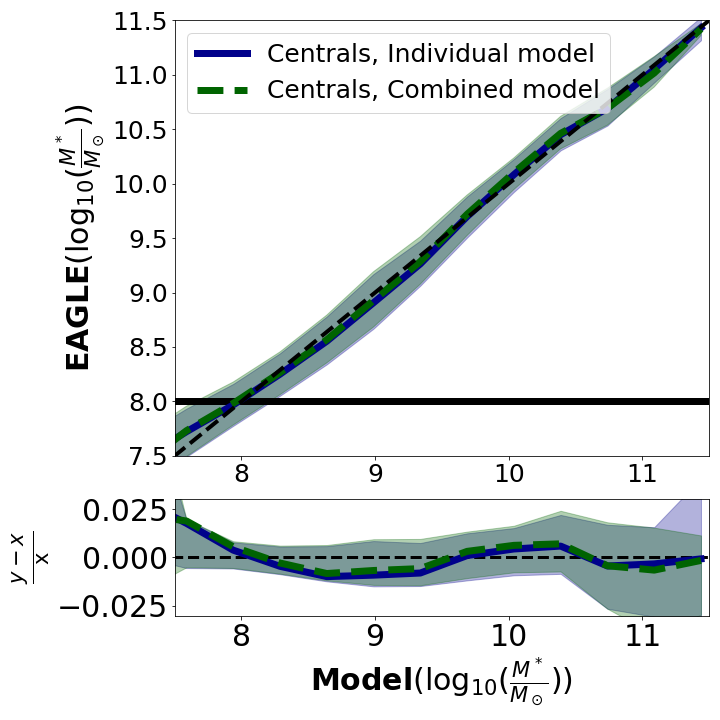}
\includegraphics[width=85mm,height=85mm]{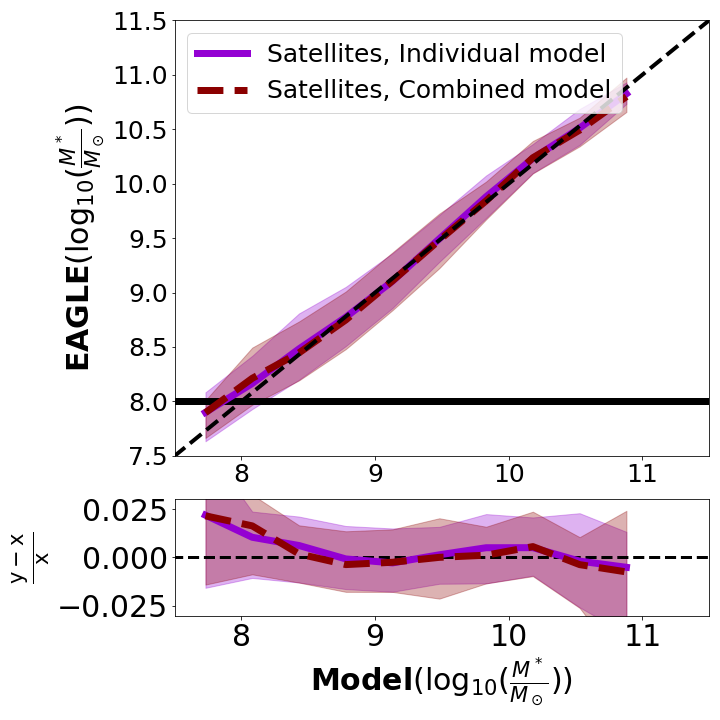}
\caption[Comparison between the stellar mass of galaxies in EAGLE and the value predicted by the model]{Comparison between the stellar masses of galaxies in EAGLE and those predicted by the models for all halos within the Holdout set. The coloured shaded areas on the top panels show the boundary encompassing $68\%$ of this holdout galaxies within bins of fixed model SM, and the solid lines are their mean values. The black dashed line corresponds to the one-to-one line. The black horizontal lines show the resolution limit of galaxies within the EAGLE simulation \citep{2015MNRAS.446..521S}. Below this line galaxies are defined by fewer particles and numerical noise starts to become an issue.  The left panel shows the result for the central halos: the solid blue line and light blue shading corresponding to the model trained on centrals alone, while the green dashed line and light green shading to the combined model, trained using centrals and satellites. The right panel is equivalent to the left panel but for satellite galaxies. The bottom panel shows the relative difference between our model prediction and EAGLE data, defined as $(y-x)/x=[\text{ EAGLE}(\log_{10}((M*/M_\odot))-\text{ Model}(\log_{10}(M*/M_\odot))]/\text{ Model}(\log_{10}(M*/M_\odot))$. It represents the relative difference between the coloured lines and shades and the one-to-one line (black dashed line) shown in the top panel.
}

\label{Model_vs_SM_plot}
\end{figure*}

Fig.~\ref{Model_vs_SM_plot} shows a comparison between the stellar masses predicted
by the models for halos in the holdout set and their actual values in EAGLE. This choice of sample enables the accuracy on the model to be assessed by considering data that was not used in training the model. The left and right panels show the results for central and satellite halos respectively. The figure shows that the mean closely follows the one-to-one relation (black dashed line) for all models above $\log_{10}(M^\star/{\rm M_\odot}) \sim 8$. The bottom panels highlight how accurate the models are, with the shaded area corresponding to an estimate of the error on the mean. The latter is computed using the central $68\%$ range of the stellar mass distribution divided by the square root of the number of galaxies in a given stellar mass bin. The mean model stellar mass is predicted to percent level accuracy for all stellar masses of interest and always within our estimate of the error on the mean.

Overall, the plot is encouraging and shows that the properties of satellites, as well as centrals, can be accurately predicted by the SRM approach. This is an important prerequisite for constructing accurate mock catalogues from dark matter simulations. We will explore the performance of the models in more detail below.

A subsidiary aim, however, is to determine whether it was necessary to explicitly distinguish between central and satellite galaxies in constructing the model. We test this by comparing the model in which central and satellite galaxies are fitted separately with one that combines all galaxies into one single model and relies on the methodology to distinguish between satellite and central galaxies only on the basis of their different formation histories. The dashed coloured lines in Fig.~\ref{Model_vs_SM_plot} show the mean stellar mass of the central (left panel) and satellite (right panel) galaxies in the holdout set when the combined model was used, i.e.\ a model that is trained on all galaxies simultaneously with no binary distinction between satellites and centrals. Those dashed coloured lines are virtually identical to the models inferred using central and satellite information alone (solid lines).

Removing this binary condition should result in an algorithm that is less dependent on the details of the SUBFIND algorithm, making results simpler to interpret.

\begin{table*}
\begin{tabular}{|c||c|c|c|c|c|c|c|}
 \hline
 &  $\log_{10} M^*/M_{\odot}$ & ${\rm lgM_{ max}}$ & $\mathbf{FC}_{20}'$ & $\mathbf{FC}_{30}'$ & $\mathbf{FC}_{50}'$ & $\mathbf{FC}_{70}'$ & $\mathbf{FC}_{90}'$ \\
\hline
$\mu$ &  8.760 & 11.13 &  3.054 &  2.481 &  1.644 &  1.034 &  0.531 \\
$\sigma$ & 0.8002 &  0.4566 & 0.8311 & 0.8786 & 0.7666 & 0.6291 & 0.5206 \\ 

\end{tabular}
\caption[Normalization parameters of DM halo variables and $M^*$]{Normalization parameters used for the stellar mass and the DM halo variables. These parameters are for the model that mixes central and satellite galaxies. The $\mu$ and $\sigma$ rows correspond to the mean and standard deviation of the variables respectively and are used in Eq.~\ref{normalisation} to standardise the range of the variables considered.}
\label{Normalisation_table}
\end{table*}

In order to compare the accuracy of the models, we use the mean square error (RMSE) statistic defined as:
\begin{equation}
\label{RMSE}
    \text{RMSE}=\sqrt{\frac{\sum_{\alpha=1}^N(M^*_{p\alpha}(C)-M^*_\alpha)^2}{N}},
\end{equation}

We find the same $\text{RMSE}$ of $0.203$ for the central galaxies in the holdout set when we predict their stellar mass with either our combined model or the model run with central galaxies only. Similarly, satellite galaxies in the holdout set have a $\text{RMSE}$ of $0.236$ in the individual model, and a $\text{RMSE}$ of $0.243$ in the combined model. This shows that a binary distinction between central and satellite galaxies does not improve significantly the accuracy of the models.

We can also look at all centrals and satellites of the individual models used together which have a $\text{RMSE}$ of $0.215$. 

This is very comparable to the $\text{RMSE}$ of the combined model which is $0.216$.

This indicates that the individual models and the combined model have comparable accuracies. Note that the combined model ends up with 21 terms while modelling satellites and centrals individually requires 14 and 12 terms respectively (hence 26 terms in total).

We would like to highlight that none of the three models shows a significant difference between the RMSE of the holdout and training sets at the third significant figure. This suggests that our methodology is robust against overfitting, as overfitting would result in a difference between the RMSE of the holdout and training set. Hence our method of selecting the hyperparameter $\lambda$ in Eq.~\ref{LASSO}, designed to avoid overfitting, works as intended.

In the rest of this work, we present our statistics using the whole data set. This is justified as we have shown that the accuracy of the models is similar for galaxies in the training set and in the holdout set. The holdout set alone is rather small (about five thousand galaxies typically), and therefore statistics like stellar mass functions or galaxy correlation functions would result with comparatively large statistical uncertainties, if the models are applied to the holdout data only.

A significant appeal of the SRM approach is that the surviving terms in Table~\ref{Results_table} have a physical interpretation. Following the discussion in \cite{icazalizaola2021sparse}, we note that there are four types of surviving parameters:
\begin{itemize}
    \item A constant, or normalisation, term.
    \item Terms that only include ${\rm lgM_{ max}}$ and no formation criteria parameter: these terms model the underlying relation between $M_{\rm max}$ and $M^*$. For central galaxies they should correspond to a model of the SMHM relation. 
    \item Terms that only include formation criteria parameters (e.g.\ $\mathbf{FC}_{50}$ and higher order combinations): these terms quantify the growth history of the halo, capturing scatter in the relation.
    \item Terms that are a product of halo mass, ${\rm lgM_{ max}}$, and formation criteria parameters: these terms model the dependence of the assembly history on the final halo mass. 
\end{itemize}

Comparing the models, we see that the constant term and the coefficients that depend only on the ${\rm lgM_{ max}}$ coefficients are similar between all three models. This reflects the similar underlying shape of the $M_{\rm max}$ and $M^*$ relation.

In the combined model, central and satellite galaxies are treated on an equal footing and their offset is captured by the more complex dependence on formation time parameters. The combined model needs 21 parameters, which are less free parameters than the combination of the two separate models, which each require 12 and 14 parameters to model centrals and satellites respectively. One noticeable difference is that the combined model relies on terms of the shape $\mathbf{FC}_i \times \mathbf{FC}_{90}$ which measure the time it takes a halo to evolve into their maximum mass with respect to the time it took them to reach a smaller percentage of that mass.

It is interesting to compare the central galaxy model with the one presented in~\cite{icazalizaola2021sparse}. It is important to stress that we do not expect identical models, since we have broadened the range of masses considered and weighted the cost function to emphasize the importance of predicting stellar masses well over the full halo mass range.
These changes resulted in a slightly simpler model.

The number of free parameters selected by the algorithm has decreased from 17 to 12. However, a close inspection of the surviving parameters of both models reveals a lot of striking similarities between the two. Many of the surviving terms are similar despite the differences in the definition of the halo mass term and, to some extent, the formation criteria definition (see Section~\ref{subsec:mass_methodology}). Here we use ${\rm lgM_{max}}$, while it was $\log(M_{200}^c)(z=0)$ in \cite{icazalizaola2021sparse}. Both models have surviving coefficients of similar amplitudes for the $constant$, the $\log(M)^x$ and the $\mathbf{FC}_{j}^\mathbf{x}$ terms (with $\mathbf{x}<3$), with a difference now that $\mathbf{FC}_{20}$ is selected instead of $\mathbf{FC}_{30}$. In summary, the main difference between both models is that the model in \cite{icazalizaola2021sparse} required more cross terms between the mass and the formation criteria parameters while now we only require one (${\rm lgM_{max}} \times \mathbf{FC}_{20}$).

One difficulty becomes apparent when comparing the models in greater detail, however. Because of the significant correlation between parameters, models of almost equivalent accuracy and complexity can vary in the final parameters chosen if these parameters are correlated. 
For example, the current central model includes strong dependencies on terms in $\mathbf{FC}_{20}$, while the model of \cite{icazalizaola2021sparse} had most terms as function of $\mathbf{FC}_{30}$.

It is difficult to decide on the significance of these differences because of the underlying correlations of $\mathbf{FC}_{20}$ and $\mathbf{FC}_{30}$. As mentioned in section~\ref{subsec:mass_methodology}, future investigations could consider methods like principal component analysis to transform our input functions into a parameter space where they are uncorrelated. However, this would lose the benefit of having a simple physical interpretation of the input parameters
and the resulting model.

\subsection{Predicting clustering and the stellar mass function}
\label{Cluster_&_SM}

\begin{figure}
\centering
\includegraphics[width=85mm,height=75mm,trim={0 0 0 0.43cm},clip]{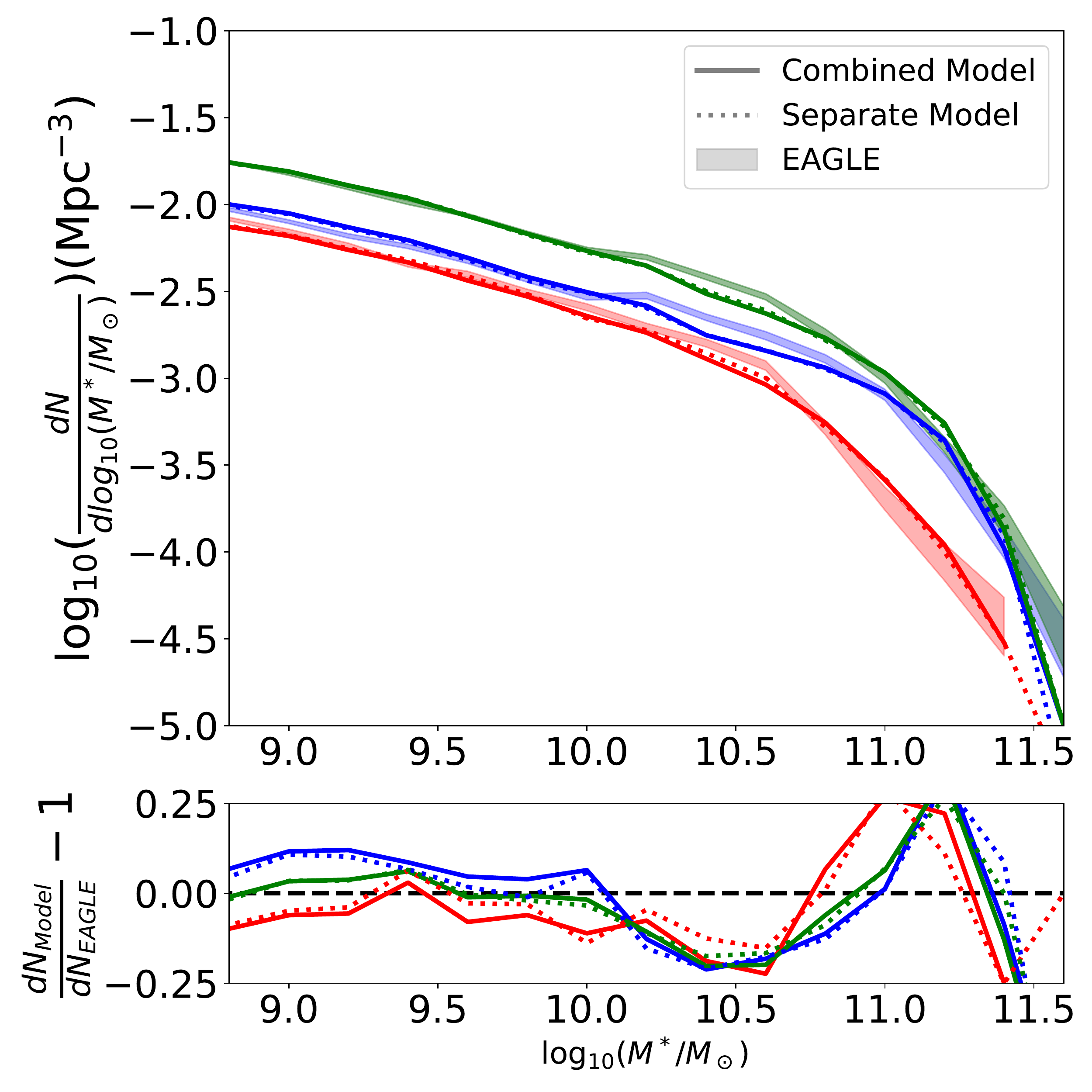}
\caption[The SMF of our models compared to that of EAGLE]{ The galaxy SMF of EAGLE, represented as the shaded areas, compared to the galaxy SMF of our models, shown as solid (combined model) and dotted (individual models) lines. The green line corresponds to combined samples of all galaxies, and the red and blue lines to the satellite and central subsets respectively. The shaded region shows the bootstrap error on the EAGLE SMF estimate. The bottom panel shows the relative difference of the model predicted SMFs compared to the EAGLE SMF, with the same line styles and colours as in the top panel. The SMFs are shown to $\log_{10}(M*/M\odot)=8.8$, the threshold below which the EAGLE galaxy sample starts to be incomplete due to our halo mass cut.}
\label{SMF}
\end{figure}

\begin{figure*}
\includegraphics[width=85mm,height=85mm]{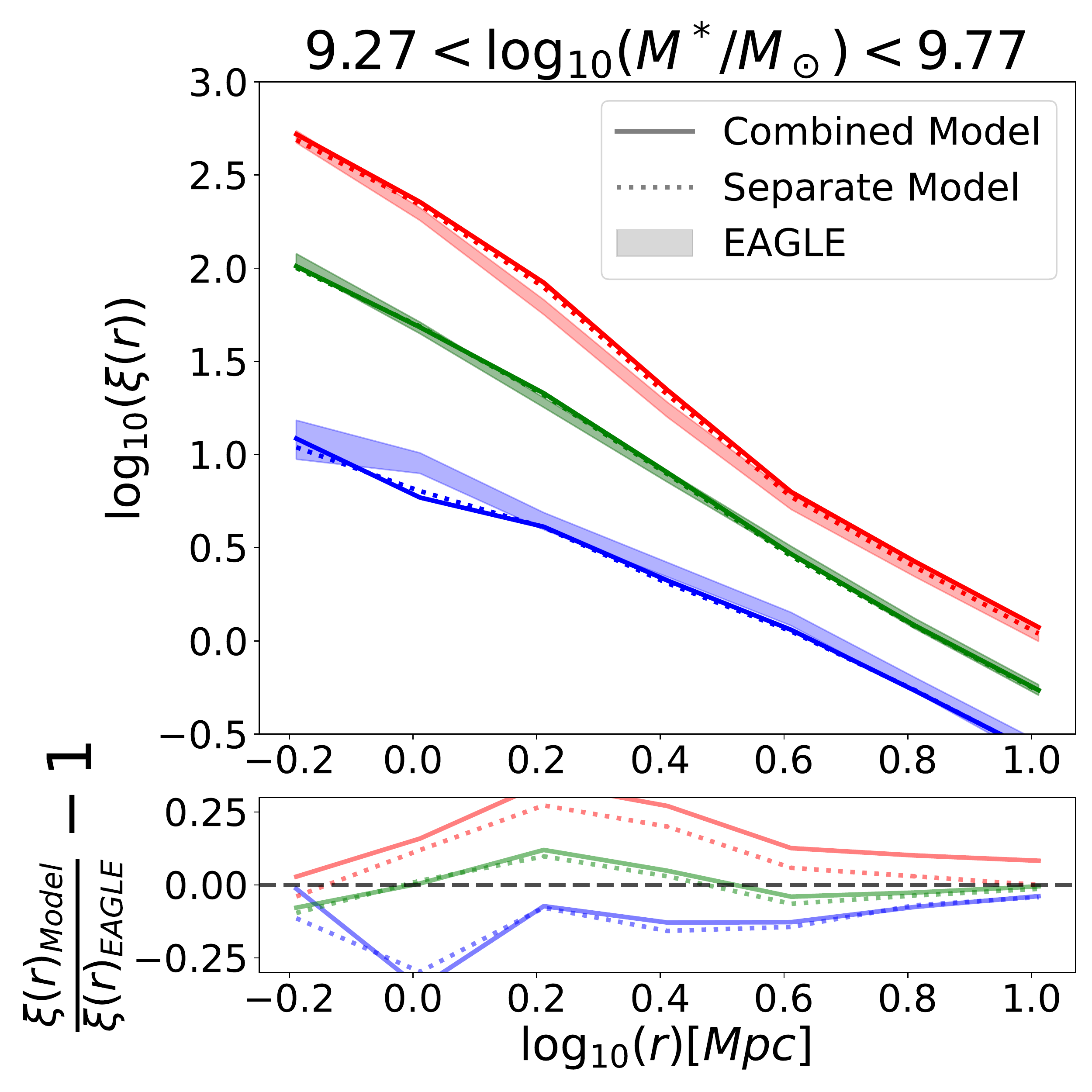}
\includegraphics[width=85mm,height=85mm]{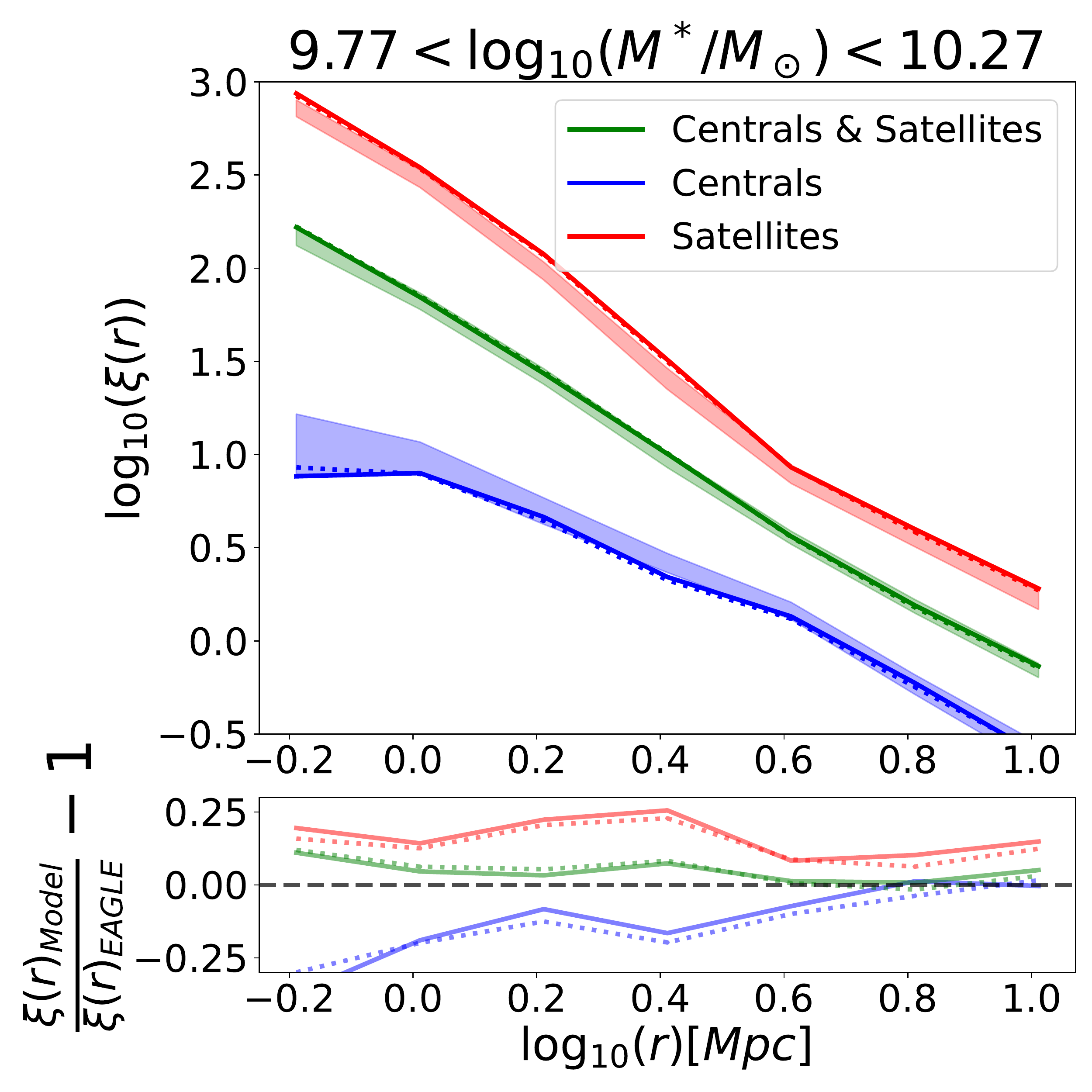}
\includegraphics[width=85mm,height=85mm]{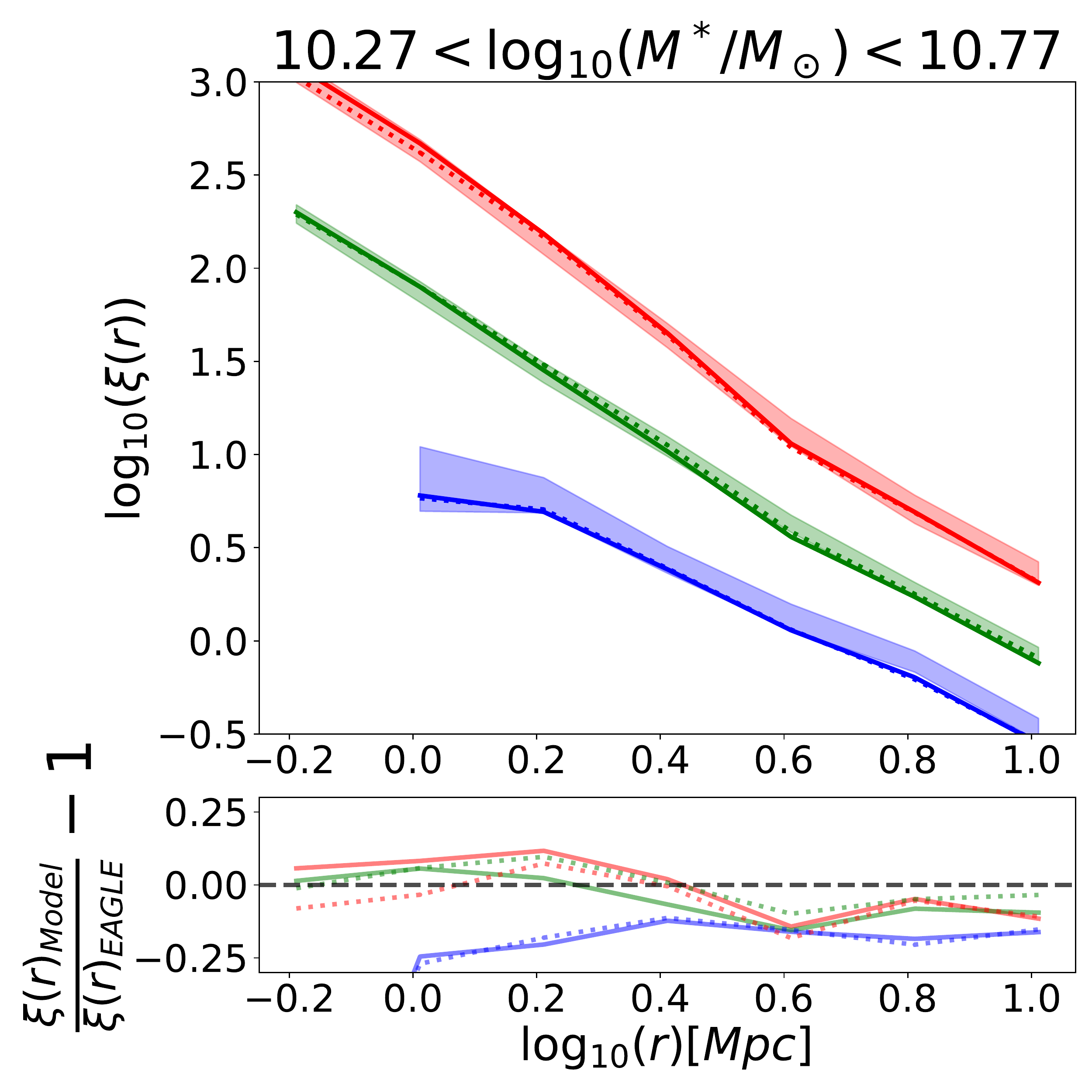}
\caption[Correlation function of EAGLE and models split into different mass bins]{Correlation function of EAGLE galaxies split into different stellar mass bins (as indicated in the title of each panel). The solid (dotted) lines show the correlation function of all galaxies in our combined (individual) models. Like in Fig.~\ref{SMF}, the colour coding refers to the galaxy sample type: all, central and satellite galaxies are in green, blue and red respectively.
The shaded area correspond to the correlation function of the corresponding EAGLE galaxies including bootstrap errors. The bottom panels show the relative difference of the model predicted correlation function compared to the EAGLE one, with the same line styles and colours as in the top panels.}
\label{Corr_function_plot}
\end{figure*}

In this section, we explore the stellar mass function (SMF) and the clustering of the stellar population generated by applying our model to the halos in our DM only simulation. We compare our resulting statistics to the ones we get from the stellar population of the full EAGLE hydrodynamical simulation. All of the statistics presented here include all halos in the DM only simulation, even those that were not matched in Section~\ref{subsec:matching}. Fig.~\ref{SMF} shows how the SMF of our models, split by galaxy type (total in green, centrals in blue and satellites in red) compares to those from the EAGLE hydrodynamical simulation. The plot shows the SMF of the combined model (solid lines), of the individual models (dotted lines) and of the EAGLE data (shaded area), with the shading indicating a bootstrap error estimate to account for sampling effects \citep{efron1979}. The different model SMFs are all comparable, as they seem to agree all similarly well with the EAGLE SMFs, with the agreement worsening somewhat for masses around $\log_{10}(M^\star/{\rm M_\odot})=10.5$, as identified already in \cite{icazalizaola2021sparse}. As we suggested in that work, one possible reason behind this disagreement is the stochasticity of certain baryonic processes which might affect the stellar mass, for example the feedback from supermassive black holes \citep{bower2017, 10.1111/j.1365-2966.2012.20879.x}. While this would be a challenging phenomenon to predict using input parameters from a DM only simulation, it should be possible to develop, in a future work, SRM models that estimate both a central value and a stochastic scatter in the predicted quantities.

In what follows we show different predictions of galaxy correlation functions and analyze how do they compare with the original statistics from EAGLE. We  emphasise that our model is not tuned to reproduce the clustering of EAGLE. Therefore any success that we may find is a consequence of correlation functions being preserved when populating the correct halos with galaxies of a given stellar mass.

Fig.~\ref{Corr_function_plot} shows the galaxy correlation functions of our models, split by the predicted galaxy stellar mass. The figure also includes the correlation function of galaxies when split by their stellar mass in the EAGLE simulation. As with Fig.~\ref{SMF} we have included an estimate of the error due to sampling effects using the bootstrap method. The correlation function of both models with central and satellites galaxies (green lines) agrees within the errors with the EAGLE correlation function. The same is true for central galaxies (blue lines). On the other hand, satellite galaxies (red lines) are slightly more strongly clustered compared to EAGLE in the lowest stellar mass bin.
There seem to be no discrepancies in the correlation functions when satellites and central galaxies are modelled together or separately. This is encouraging as it implies that the binary distinction between central and satellite galaxies becomes unnecessary to model the overall correlation function using our prescription.

One of the advantages of our methodology over standard machine learning techniques is the fact that our solution is expressed as a simple equation of state with 21 free parameters fitted by the algorithm. This is important as the model coefficients can be modified so that other data sets (different from EAGLE) can be fit. This would be needed when for example one wants to populate DM only simulations with EAGLE informed physical processes to create mocks that mimic observational data set. This could not be achievable by a more complex {\it black box} model.

\subsection{Comparison with other models}
\label{other methods}

\subsubsection{Comparison with SHAM}
We have stated that we are interested in using our methodology as an alternative for populating halos in DM only simulations with galaxies. To test if our methodology is adequate, we first need to compare our accuracy to that obtained from standard methods like sub-halo abundance matching (SHAM) \citep[e.g.][]{2004MNRAS.353..189V,2006ApJ...647..201C}, that makes a one-to-one matching between halos and galaxies, based on a property that correlates with the stellar mass. More recent implementations of SHAM add some stochasticity to the methodology to account for the scatter in the correlation \citep[e.g.][]{2010ApJ...717..379B,2014}. Therefore, regular SHAM implementations produce models that depend on only one free parameter and one subhalo property, which makes them simpler than our SRM models that consider six halo properties and fit several free parameters.

In what follows, we compare the correlation function from our combined model to the one presented by \cite{2016MNRAS.460.3100C}. They used a SHAM methodology to populate galaxies in the EAGLE simulation by studying the relation between the stellar mass of a galaxy and the maximum circular velocity of a halo once it reaches equilibrium after a merger.

\begin{figure*}
\includegraphics[width=85mm,height=85mm]{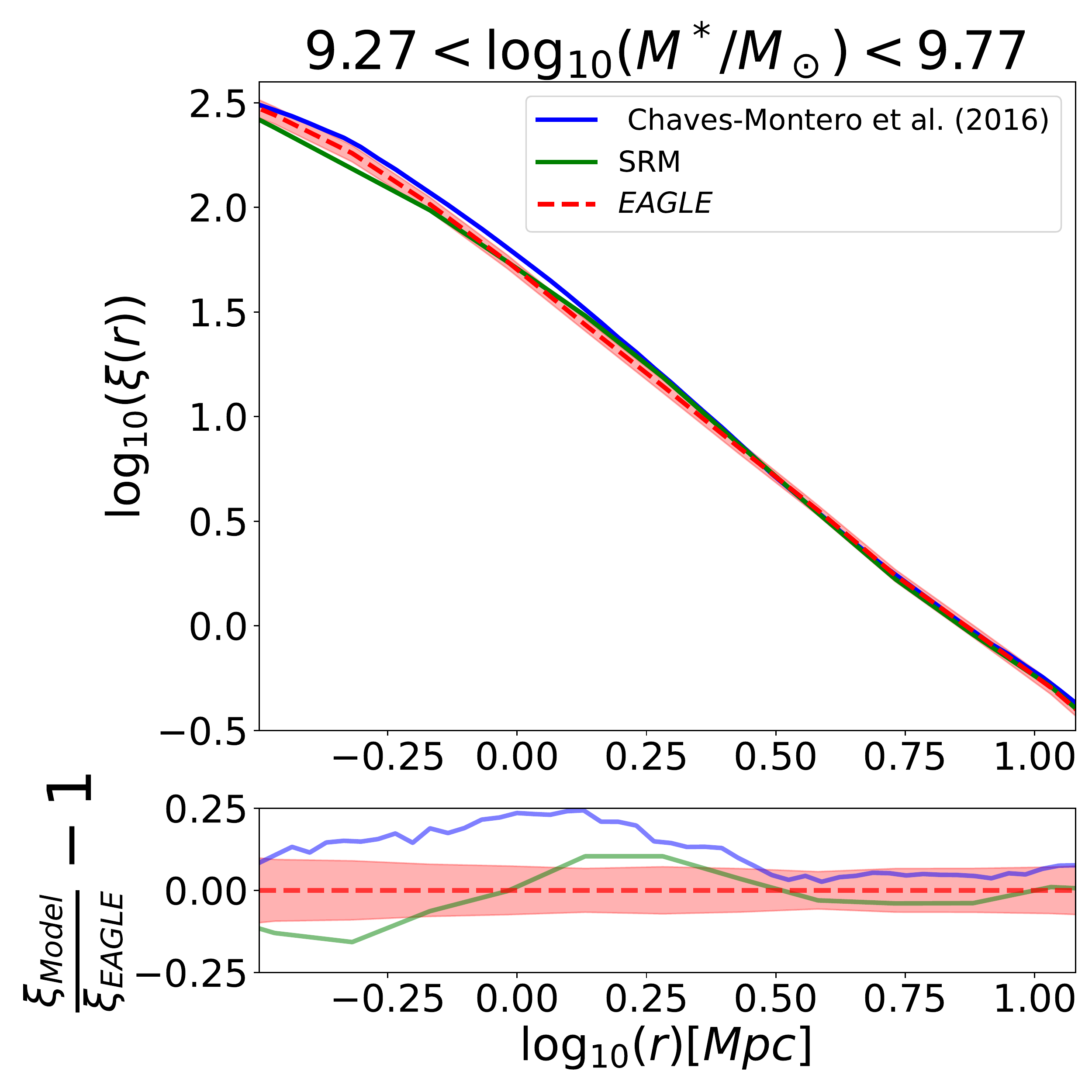}
\includegraphics[width=85mm,height=85mm]{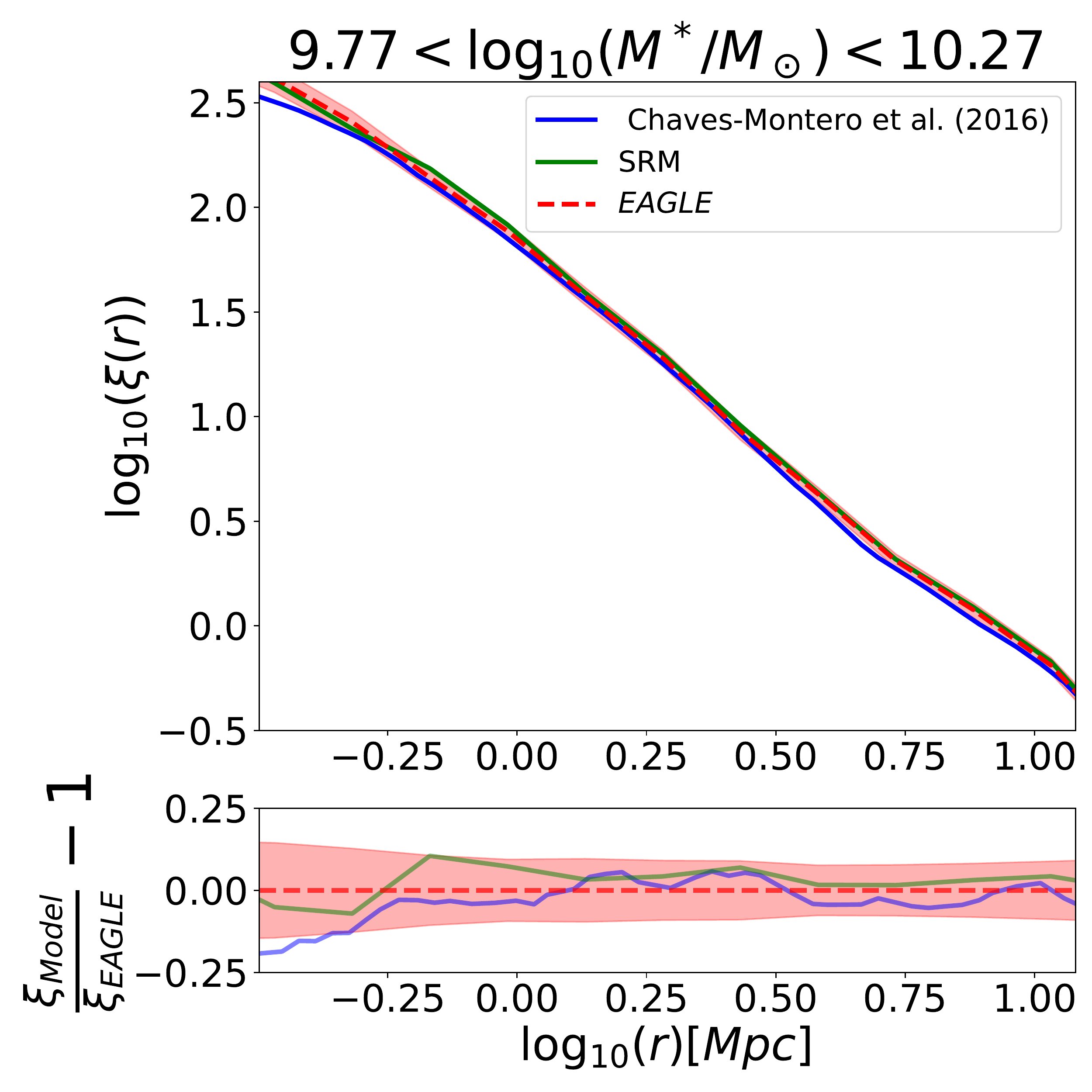}
\caption[Correlation function of our model compared to SHAM]{Correlation functions of the combined model produced with our SRM method (green solid lines), of the SHAM results presented in \cite{2016MNRAS.460.3100C} (blue solid lines), and of the EAGLE hydrodynamical simulation (red dashed line and shading). The correlation functions are computed for galaxies in the stellar mass bins indicated in the title of each panel. The shading correspond to bootstrap errors. The bottom panels show the relative difference of the model predicted correlation functions compared to the EAGLE ones, with the same line styles and colours as in the top panels.}
\label{CM_comparison}
\end{figure*}

Fig.~\ref{CM_comparison} shows how our correlation functions (blue lines) compare to the ones from \cite{2016MNRAS.460.3100C}. The right panel shows that for larger stellar masses, both methods agree with EAGLE within the bootstrap errors,  while they provide reasonable accuracy in recovering the correlation function for smaller stellar masses (as shown by the left panel). However \cite{2016MNRAS.460.3100C} seems to struggle to recover the EAGLE correlation function on the smaller scales. They report differences of $20\%$ to $30\%$, as confirmed in the bottom left panel of Fig.~\ref{CM_comparison}. Our SRM model shows a slight improvement on these smaller scales and agrees better with the EAGLE correlation function. For stellar masses larger than the ones shown in Fig.~\ref{CM_comparison} we continue to agree with the EAGLE simulation within errors.

\subsubsection{Comparison with Machine Learning Tree methods}
\label{Lovell_comparison}
We have stated that our goal is to develop an explainable machine learning methodology. However, for this to be of use we need to make sure that the accuracy of our model is comparable to that of more established machine learning (ML) methods. With this in mind, in what follows we compare our model with the ML model presented in \cite{lovell2021machine}, which uses extremely randomized trees (ERT) \citep{ERT_paper} to model galaxy properties from EAGLE halo information. ERT methods are emerging as a popular and highly accurate ML method to model the relations between galaxies and host halos \citep[e.g.][]{10.1093/mnras/stv2310,10.1093/mnras/stz2304}. 

The model of \cite{lovell2021machine} is trained using data from the EAGLE and the C-EAGLE simulations \citep{10.1093/mnras/stx1647,10.1093/mnras/stx1403}. The latter is a set of zoom-in hydrodynamical simulations of massive galaxy clusters. The calibration of C-EAGLE is slightly different from the standard EAGLE one, with changes in the values of the parameters determining the AGN feedback and the black hole accretion rates. This new parametrization is usually referred to as AGNdT9 \citep{2015MNRAS.446..521S}. The EAGLE data used in \cite{lovell2021machine} comes from a smaller box of 50 comoving Mpc, that has the same resolution and cosmology as the standard 100 Mpc box, but uses the AGNdT9 parametrisation of C-EAGLE. 

We decided to compare with the model of \cite{lovell2021machine} as it was also constructed using the EAGLE simulation and therefore it shares the same cosmology and resolution and was built using the same algorithm that our data, which makes a direct comparison of the models more straightforward. ML methods that have been trained on other simulations might have differences in the accuracy of the models that could be a consequence of the training data and not of the methodology itself.

\cite{lovell2021machine} uses either eight or twelve properties of the host DM halos to model the stellar properties of galaxies (depending on the specific model). Hence the number of input parameters they consider is comparable to our work, as we use six halo properties. Their properties include information that parameterize the host halo mass at $z=0$, like the total mass of the halo $M_{FoF}$, and properties that are more correlated with the assembly history, like $V_{max}$ or the radius at which $V_{max}$ is reached. On the other hand our formation criteria parameters contain a more direct parametrisation of the assembly history.

\begin{figure*}
\includegraphics[width=85mm,height=85mm]{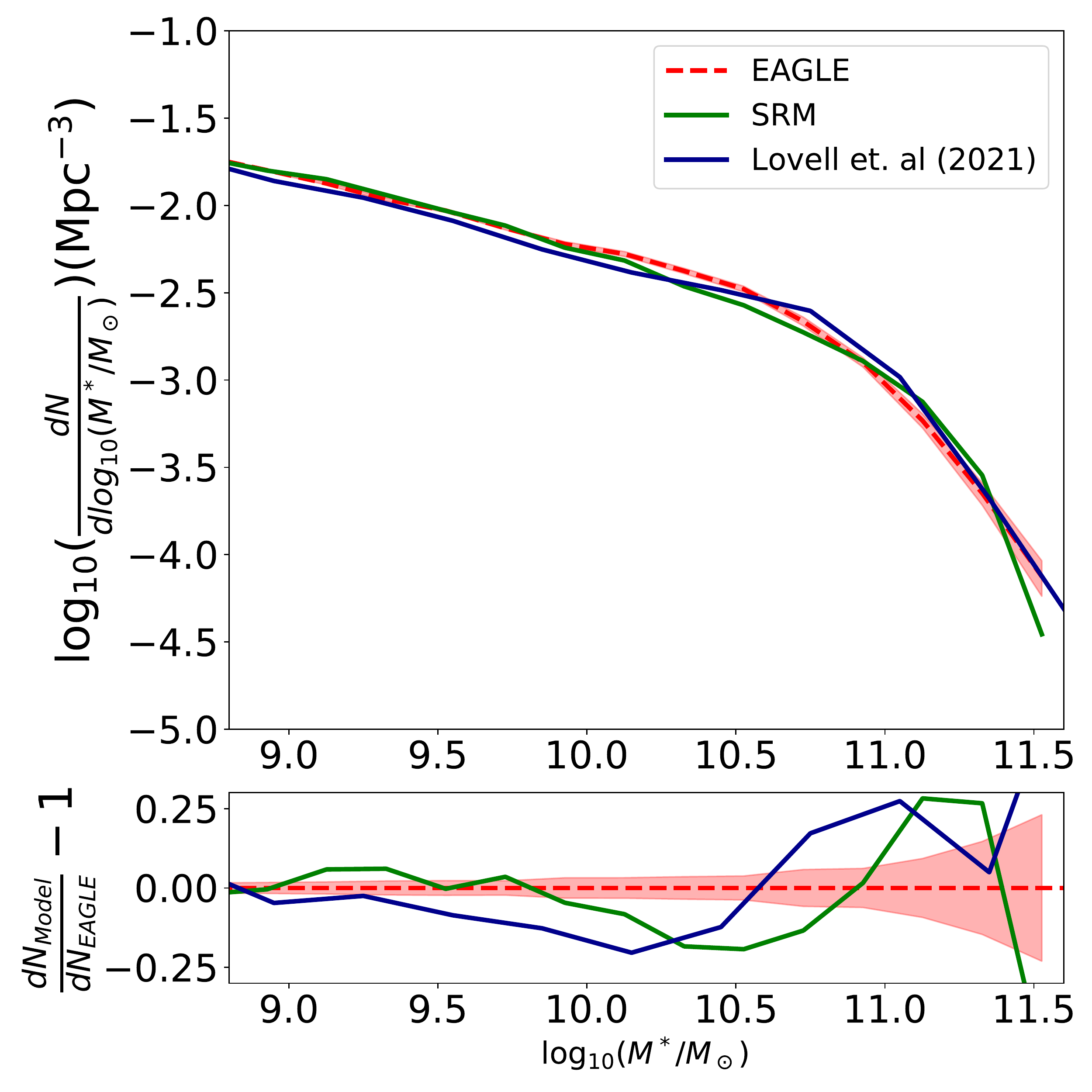}
\includegraphics[width=85mm,height=85mm]{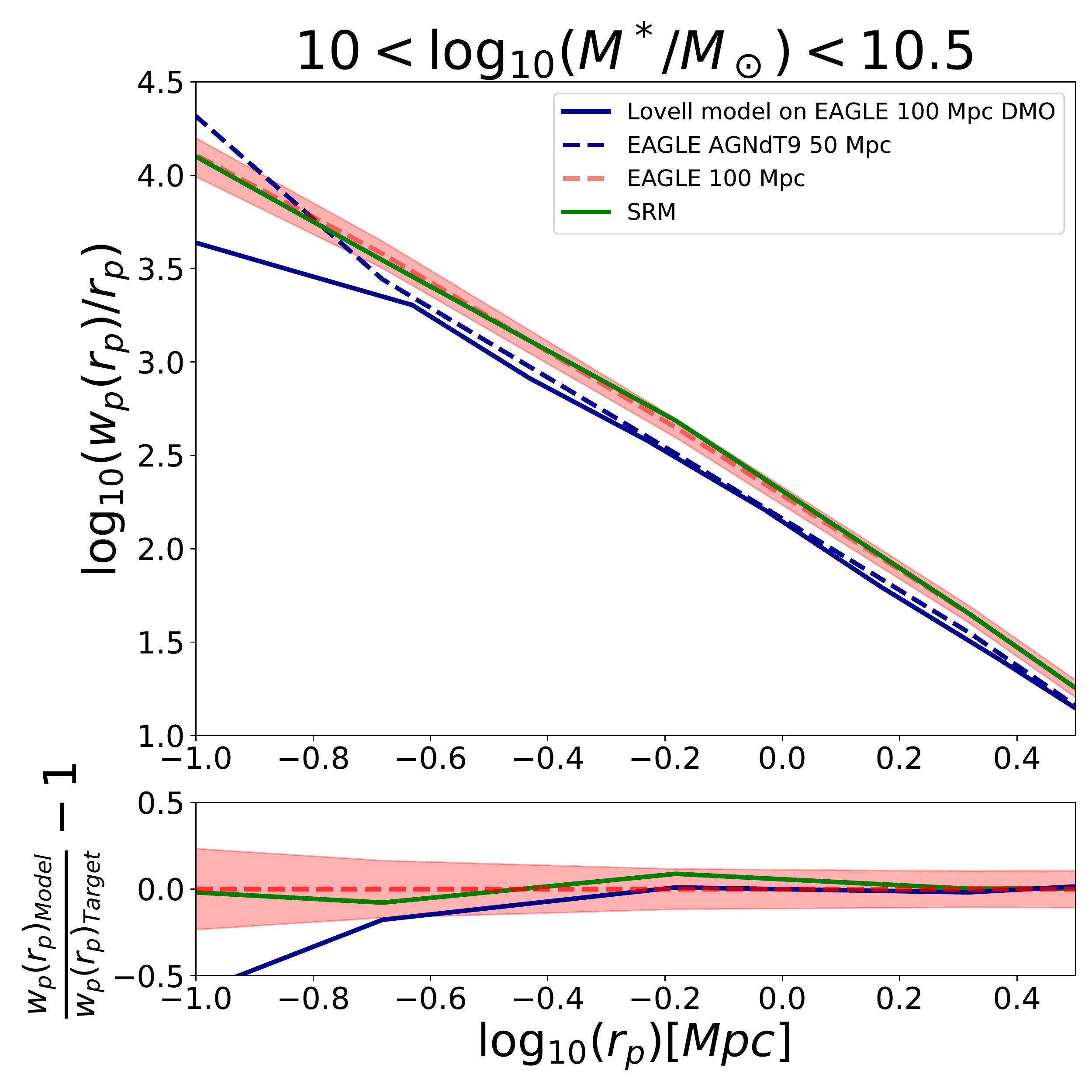}
\caption[Correlation function of our model compared to Lovell]{{\bf Left panel:} the SMF predicted by the ERT method of \cite{lovell2021machine} (blue line) and our SRM model (green line) when applied to halos within the EAGLE DMO simulation box. {\bf Right panel:} The projected correlation functions predicted by \cite{lovell2021machine} (blue solid line) and by our SRM model (green solid line) when applied to halos within the EAGLE 100 Mpc DMO simulation, using the stellar mass bin of \cite{lovell2021machine}. In both top panels, the red dashed lines (and shading) show the corresponding statistics (and bootstrap errors) measured directly from the EAGLE hydrodynamical simulation. The blue dashed line of the right panel shows the projected correlation function of the EAGLE simulation built with the AGNdT9 parameterization. The bottom left panel show the relative difference of the model predicted SMF w.r.t.\ the EAGLE 100 Mpc box, with the same line styles and colours as in the corresponding top panels.  The green line in the bottom right panel shows the relative difference of the projected correlation functions of our SRM model w.r.t.\ the one from the EAGLE 100 Mpc box, while the blue line is the relative difference of the projected correlation function from \cite{lovell2021machine} model w.r.t.\ the one from the EAGLE AGNdt9 50 Mpc box. We note that \cite{lovell2021machine} model is trained on data from both the C-EAGLE and EAGLE AGNdt9 simulations, and therefore it is less likely to reproduce the SMF of EAGLE as accurately as our model that is trained solely on
EAGLE. See the main text for a more detailed discussion.}
\label{SMF_Lovell}
\end{figure*}

The top left panel of Fig.~\ref{SMF_Lovell} shows how the SMFs from our model and from \cite{lovell2021machine} compare to the one from the EAGLE hydrodynamical simulation.
The bottom left panel of Fig.~\ref{SMF_Lovell} shows the relative difference of the model SMFs w.r.t.\ to the EAGLE 100~Mpc. 
We note that both models have comparable accuracy, with our SRM model being slightly more accurate for stellar masses between $\sim 10^9$ and $\sim 10^{10}$~M$_\odot$. 
However, we should emphasize that given the  \cite{lovell2021machine} model is trained on a combination of C-EAGLE and AGNdt9 data, it is less likely to reproduce the SMF of EAGLE as accurately as a model that is trained solely on EAGLE, like ours. 

\cite{McAlpine_2016} show that the SMF of the AGNdt9 simulation agrees well with the one from the larger EAGLE 100~Mpc simulation, with both SMFs being identical in all but the larger stellar mass bins where AGNdt9 lacks volume to be representative, which is precisely what the C-EAGLE data used by \cite{lovell2021machine} compensates for. Therefore it is reasonable to compare both models with the SMF of the EAGLE 100 Mpc box, bearing in mind those limitations.

The right panel of Fig.~\ref{SMF_Lovell} shows the projected correlation function ($w_p(r_p)$) \footnote{The projected correlation function \citep{1983ApJ...267..465D} is defined as: $w_p(r_p)=2\int_{-\infty}^{\infty}\xi(r_p,\pi)d\pi$,  where $r_p$ and $\pi$ are the components of $r$ perpendicular and parallel to the line of sight respectively.} for a stellar mass selected sample as defined by the panel title.
The clustering of the 50~Mpc box built with the AGNdT9 parametrisation is slightly different from the one built with the standard 100~Mpc box, as shown by the two dashed lines in the right panel of Fig.~\ref{SMF_Lovell}. The AGNdT9 simulation (along with C-EAGLE data) was used to build \cite{lovell2021machine} model and therefore the correlation function of the model applied to the DMO should be compared with the correlation function of the AGNdT9, which is why the ratio of the bottom panel of the right plot is done w.r.t.\ the the EAGLE AGNdT9 50~Mpc box.

 The two solid lines in the right panel of Fig.~\ref{SMF_Lovell} correspond to two different models, as indicated by the key. We note that the projected correlation functions of our SRM model agrees well with the one from EAGLE: on all scales considered, the line from our SRM model is within the bootstrap errors of the EAGLE sample. 
 
Similarly, the clustering of Lovell's model applied to the DMO simulation agrees well with that of the EAGLE AGNdT9 simulation used to build the model. The accuracy with which this model reproduces the projected correlation function is similar to the one from our model in all but the smallest scales. This is clear from the bottom right panel of Fig.~\ref{SMF_Lovell} where the relative clustering difference of the \cite{lovell2021machine} model with respect to that of the EAGLE AGNdT9 sample is shown by the blue solid line. As the \cite{lovell2021machine} model was tuned to a combination of C-EAGLE and AGNdT9 data it is not straightforward to make a direct comparison with the clustering of their training data, a comparison to the clustering of the AGNdT9 simulation is therefore the best alternative.

We have shown that the projected correlation function and the SMF resulting from our SRM methods are comparable to the ones obtained by \cite{lovell2021machine} using ERT methods. As we have stated, the comparison between \cite{lovell2021machine} and our model cannot be done fully accurately, as they use data from the C-EAGLE simulation to build their models. Nevertheless, we consider the fact that the models seem to have a similar level of accuracy as an encouraging result, especially as ERT methods are designed to be accurate and cost-efficient \citep{ERT_paper}. Unlike our SRM method, explainability is not an aim within the design philosophy of ERT models.

\subsection{Models with additional halo properties}
\label{OtherParameters}

The parametrization of halo properties presented in Section~\ref{subsec:mass_methodology} is different from the parameters selected by other machine learning methods. For example, \cite{lovell2021machine} uses exclusively properties at z=0 to build a model with an accuracy comparable to ours. \cite{lovell2021machine} finds that the maximum circular velocity ($V_{\rm max}$), the half mass ratio ($R_{1/2}$), the mass of the halo at $z=0$ ($M_0$), and the potential energy of the halo ($E_p$) are the parameters that have significant contributions to their stellar mass model \citep[see Figure 11 of][]{lovell2021machine}. 

In this section we explore whether some of these parameters could improve our baseline model, as presented in section~\ref{sec:results}. This is not a trivial question, as some of these parameters, like $V_{\rm max}$ and $R_{1/2}$, might be useful in other machine learning models as they are a better tracer of the inner part of the halo than $M_{\rm max}$. However, the halo evolution is already well tracked in our model by our parametrization of the halo evolution with the $FC_i$ parameters. 

\begin{figure*}
\includegraphics[width=85mm,height=85mm]{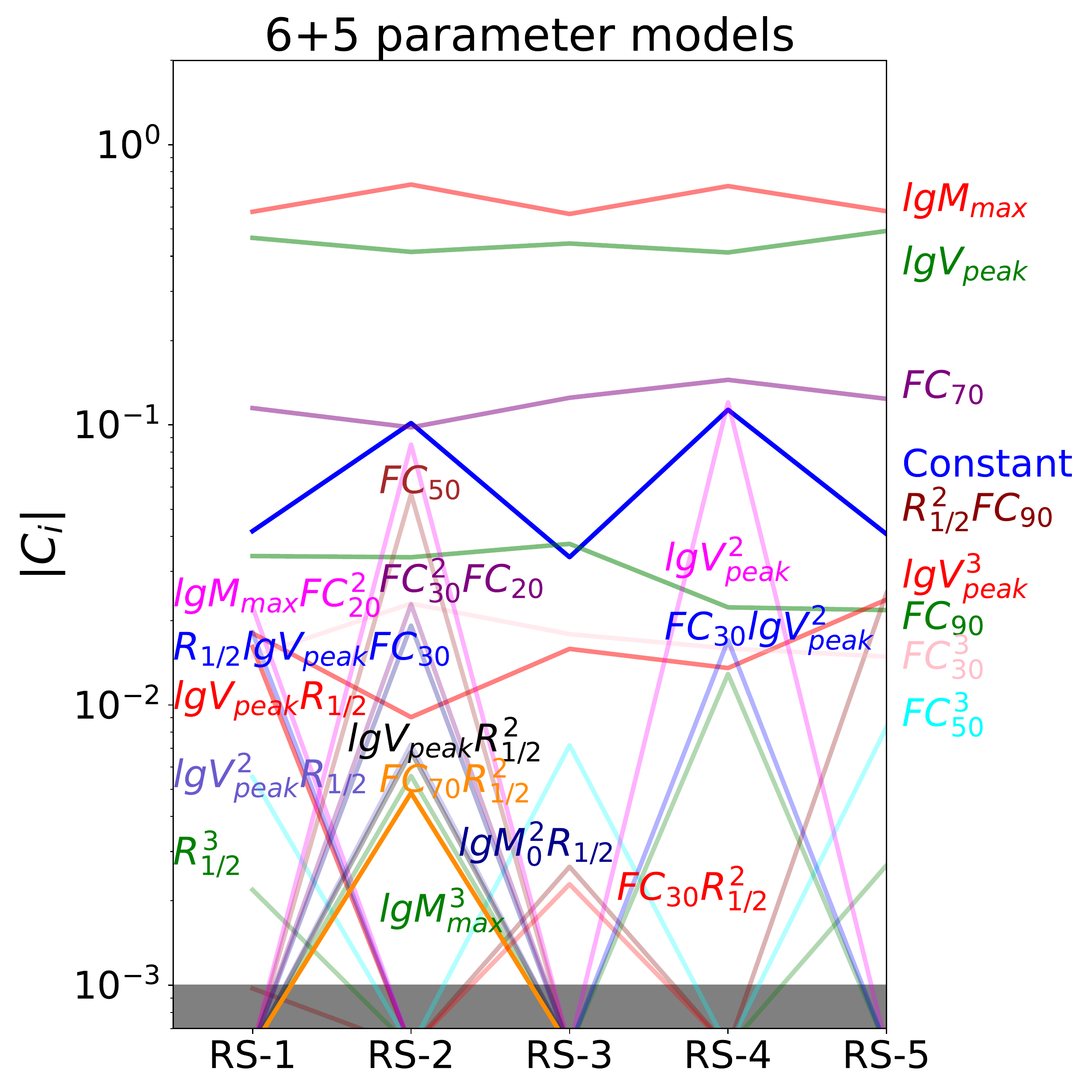}
\includegraphics[width=85mm,height=85mm]{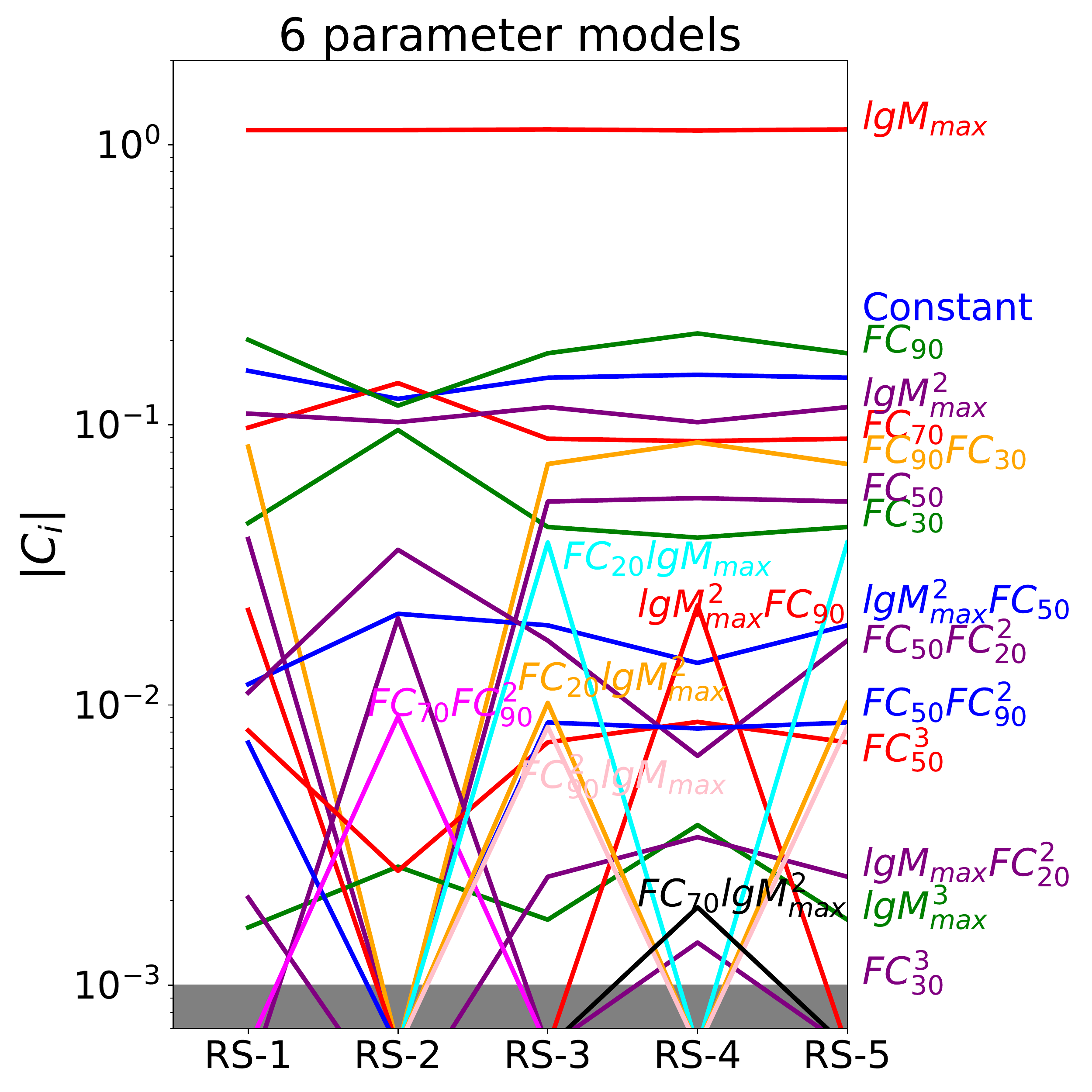}
\caption{The left panel shows the absolute values of all selected coefficients for each of the five statistically equivalent 6+5 parameter models (RS-1 to RS-5) trained and validated on a sample of 4,000 galaxies. The right panel shows similar information using the same data set, but for our standard 6 parameter model.
The coloured labels at the right of each plot correspond to parameters that were used in at least three of the five models, while the labels inside of the plot correspond to parameters used only in one or two of the models. The grey shading shows the threshold below which coefficients are discarded by a given model.}
\label{Diff_param_models}
\end{figure*}

Another issue when including additional parameters is that some might be strongly correlated with each other. In Appendix~\ref{Correlated Parameters}, we show that parameters like $M_0$ and $E_p$ will provide essentially the same information to our models. Including highly correlated parameters in our current implementation could reduce the explainability of the model, as coefficients corresponding to different polynomial terms of correlated parameters can have different physical interpretations while modeling the same underlying behaviour. 

In SRM, the standard approach for dealing with extra variables that one does not know if they could improve a model or not is to add them as free parameters and to see if the algorithm discards them by itself. This is one of the original design philosophies behind these methodologies, as discussed in \cite{Brunton3932}. 
Hence, we run our methodology using our regular six halo parameters to which we add five extra parameters: four free parameters defined at $z=0$ and suggested by \cite{lovell2021machine} ($V_{\rm max}$, $R_{1/2}$, $\log_{10}(M_0)$ and $\log_{10}(E_p)$) and a fifth parameter $V_{\rm peak}$, defined in Section~\ref{subsec:mass_methodology}.
Throughout the rest of this work we use the following unitless parameters:
\begin{equation}\label{Vpeak_definition}
\begin{aligned}
   & {\rm lgV_{peak}} =\log_{10}(V_{\rm peak}/({\rm km/s})) \\
   & {\rm lgV_{max}} =\log_{10}(V_{\rm max}/({\rm km/s}))\\
   & {\rm lgE_p} =\log_{10}(E_p/{ {\rm (M_\odot(km / s)^2}}))\\
   & {\rm lgM_0} =\log_{10}(M_0/{\rm M_\odot}).
\end{aligned}
\end{equation}

Several of these new halo properties are correlated with each other, as shown in Appendix~\ref{Correlated Parameters}. As discussed in Section~4.3 of \cite{icazalizaola2021sparse}, correlated parameters have the effect of generating multiple local minima, resulting in a highly non-convex configuration space to be explored. In such spaces, our implementation of the minimization algorithm struggles to find the global minimum, as it spends time exploring unstable local minima.
This, in turn, has the net effect of building models with slight variations in the surviving coefficients, which depend on the starting point of the minimization and on the specific selection of galaxies in the holdout set.
To address this limitation, we run our methodology five times, using the same initial set of galaxies, but modifying the random seed so that the subset of galaxies selected for the holdout set and the starting points for the minimization algorithm change. These five runs provide an idea of the average model that can be built with this new configuration.
Finally, by adding 5 new parameters to the model, the number of coefficients to minimize over goes from 84 to 364, which increases significantly the dimensionality of the problem. 

These three observations, i.e.\ correlated parameters, the need for statistically equivalent runs and the larger dimensionality of the problem, have the net effect of increasing significantly the computational cost of running our algorithm. As a consequence, we make the compromise of using only 4,000 randomly selected galaxies to run our models on (as opposed to the nominal 35,456 galaxies), as this keeps the overall computational running costs manageable. In parallel, we run another set of five models that uses our standard configuration of six parameters from Section~\ref{sec:results}, but built with the same 4,000 galaxies and random seeds as these new models. These five models correspond to our baseline models throughout this subsection, and we refer to them as our 6 parameter models. In the rest of this section these models are contrasted with their equivalent models built with extra parameters but the same random seeds, to which we refer to as our 6+5 parameter models.

Before analyzing the subset of parameters selected by the algorithm for the new 6+5 parameters models, we show that these models are as accurate as the models run with the nominal 6 parameter configuration. All five 6+5 parameters models have a RMSE between 0.22 and 0.23, which is comparable to within the uncertainty of the model fitting to the RMSE of the corresponding nominal 6 parameter models, which is between 0.21 and 0.22. These values are also comparable with our final model from Section~\ref{sec:results}, that has an RMSE of 0.22 when estimated with the set of 4,000 galaxies used in this section. We note that all five runs of the 6+5 parameter models choose a similar number of surviving coefficients, with two runs selecting 13 and 15 coefficients each, while the other three runs all selecting 10. This is in agreement with the variance on the methodology due to variations in the holdout set selection found in \cite{icazalizaola2021sparse}. We find no correlation between the number of surviving coefficients and the RMSE of the models.

Fig.~\ref{Diff_param_models} shows the values of each of the selected coefficients for our five new models using both our new configuration with 6+5 parameters (left) and the standard configuration with 6 parameters (right). 

Via the SRM methodology, most models will have a subset of their allowed parameters discarded (when none of the coefficients associated with these parameters are chosen by the algorithm). In our case, the five runs of the 6+5 parameter model end up keeping between 5 and 9 parameters. 
The differences in the number of surviving parameters between the runs show how correlated these parameters are with each other, making them somewhat interchangeable. As shown in the left panel of Fig.~\ref{Diff_param_models}, the only new parameter selected by all five runs is $\rm lgV_{peak}$, while four out of the 6 standard model parameters ($\rm lgM_{max}$, $FC_{30}$, $FC_{70}$ and $FC_{90}$) are kept in each resulting new model.

We note that all of the 6 parameter models keep all input parameters without discarding any. This suggests that the information contained inside the parameters used in our standard configuration is more unique that the one from 6+5 parameter models used in this section. This is further discussed in Appendix~\ref{Correlated Parameters}, where we show how most of the new parameters included are strongly correlated with each other and with $\rm lgM_{max}$, while the correlations between the formation criteria parameters are comparably weaker and hence contain more specific information. The fact that the models that start with 11 free parameters require in some cases a large number of parameters to reach an accuracy similar to the model in Section~\ref{sec:results} suggests that the new parameters did not contain much additional (if any) new information that was not already present in our initial model.

The five runs with our standard 6 parameter model selected between 13 and 18 coefficients, less than the 21 coefficients of our final model from section~\ref{sec:results}. The differences in selected coefficients is due to these new models being built with less data. Given that the SRM method is very strict about avoiding over fitting, it becomes harder to justify a larger set of coefficients. As a test of this, we ran another set of five models using our standard six parameters but using 12,000 galaxies (3 times more than the data in this section and 3 times less than the nominal set) and we found that the selected models use between 16 and 19 coefficients.

From the left panel of Fig.~\ref{Diff_param_models}, we note that the five 6+5 parameter models use between 10 and 15 coefficients, which is slightly less that the 13 to 18 coefficients of the 6 parameter models. This suggests that, while the new parameters might not necessarily contain much new information required to model the SMHM relation, they might be more efficient at compressing the relevant information.

While the 6+5 parameter models select on average less coefficients, it is significantly less consistent in the subset of coefficients selected by any particular model. This is shown by the fact that only 9 coefficients are selected in at least three models, while another 14 coefficients are selected once or twice only. This is in contrast with the 6 parameters models, where most coefficients are present in at least three models and only 6 out of 21 coefficients are selected once or twice. In fact, we note that even if the 6+5 parameter model uses fewer coefficients per model, the number of coefficients selected by at least one model (23 coefficients) is comparable to that from the 6 parameter model (21 coefficients). This shows how the inclusion of correlated parameters increases the stochasticity of the method, which in turn complicates the interpretation of the resulting models. 

\begin{figure}
\includegraphics[width=85mm,height=85mm,trim={0 0 0 0.43cm},clip]{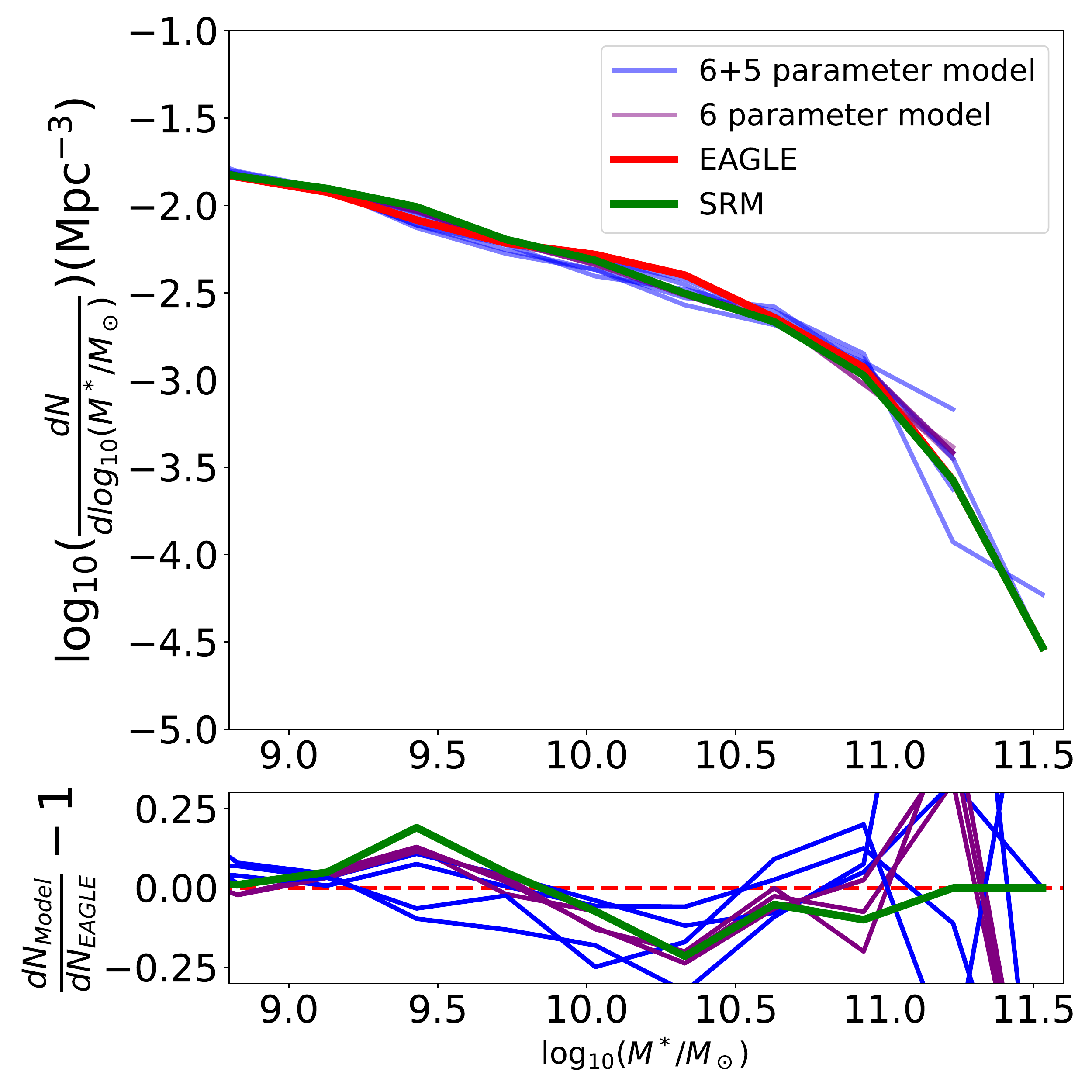}
\caption{The SMF as predicted by the five new models (blue lines) that are built by adding five new parameters to our method. The purple lines show the SMF of their corresponding SRM models built with our standard configuration of six parameters. We also include the SMF predicted by our final model from Section~\ref{sec:results} (green line) when applied to the same subset of 4,000 galaxies. The EAGLE SMF of this subset is shown as the red line. The bottom panel shows the ratio of each model predicted SMF to the EAGLE SMF, with all models predicting the SMF to a similar level of accuracy.}
\label{Diff_param_smf}
\end{figure}

Out of the 11 parameters, all 6+5 parameter models select linear contributions from $\rm lgM_{ max}$, $\rm lgV_{peak}$, $FC_{70}$ and $FC_{90}$ and cubic contributions from $\rm lgV_{peak}$ and $FC_{30}$, in order of decreasing linear coefficient value. Almost all models select some contribution from $R_{1/2}$ and $FC_{50}$ as well, except for one model, RS-4 in the left panel of Fig.~\ref{Diff_param_models}. That latter model has a comparable RMSE to the other models and requires ten coefficients, the same number as two other models. 

This highlights the difficulties of using our current SRM implementation on spaces with highly correlated parameters: RS-4 has an accuracy (RMSE) and simplicity (number of parameters) that are equivalent to those of the other four models (within the variations of the methodology due to different holdout sets). Therefore it is neither better nor worse than the other models within the standards that we designed our models to meet. However, due to parameters being strongly correlated and sharing similar information, we see that this model requires two less free parameters and therefore would have a simpler physical interpretation than the others.

As shown in Section~\ref{Correlated Parameters}, out of all new input parameters, $R_{1/2}$ is the one that is the least correlated with the rest of the new parameters. This suggests that the $R_{1/2}$ information provided to the model is possibly more unique than that from some of the other new parameters, which might explain why four new models had some contributions from the $R_{1/2}$ parameter.

We note that all models discard contributions from $\rm lgV_{max}$ and $\rm lgE_p$ up to order 3, and that only one model (RS-2 in the left panel of Fig.~\ref{Diff_param_models}) includes a very minor contribution from $\rm  lgM_0$ in the form of the coefficient ${\rm  lgM_0^2} \, R_{1/2}$. This suggests that the contribution to the accuracy of the model after including any of these three parameters is negligible and that none of these parameters contributed additional information that was not already provided without them.

$FC_{20}$ is only selected by two models, RS-1 and RS-2 in the left panel of Fig.~\ref{Diff_param_models}. Those models are the two that have the largest number of coefficients, which suggest that the information that was previously provided by $FC_{20}$ to our model is also contained in some of the new set of parameters.

In summary and as stated already, all five runs of the 6+5 parameter model selected contributions from $\rm lgM_{ max}$, $\rm lgV_{peak}$, $FC_{70}$, $FC_{30}$, $FC_{90}$. Of these parameters, $\rm lgV_{peak}$ is the only one that is not within our original set of parameters. 
Given that the five new models do not seem to be more accurate than the model presented in Section~\ref{sec:results}, the contribution provided by $\rm lgV_{peak}$ could also be obtained by a combination of the $FC_i$ parameters within our original model, as shown in Appendix~\ref{Vpeak}. However, the fact that these new models require in general less coefficients seems to indicate that including $\rm lgV_{peak}$ is an efficient way of compressing some of the information contained in our $FC_i$ parameters.

Fig.~\ref{Diff_param_smf} shows the SMFs of the five 6+5 parameter models (blue) and the complementary 6 parameter models (purple). These SMFs are built using a subset of 4,000 galaxies and we account for this sampling in the SMF estimates. The bottom panel indicates that all models have a similar accuracy (to within 25\% typically) when predicting the SMF of EAGLE. This suggests that including the extra parameters does not improve our ability to reproduce the stellar mass distribution. In addition, while the corresponding 6 parameter models do not have better accuracy than their 11 parameter counterparts, they seem to be far more consistent with each other. This can be seen by the purple lines being more similar to each other than the blue ones in the bottom panel of Fig.~\ref{Diff_param_smf}. This is due to the 6 parameter models being more consistent in their selection of surviving coefficients, as shown in the right hand panel of Fig.~\ref{Diff_param_models}.

As mentioned already, these models are trained with a small subset of the full data (4,000 galaxies as opposed to 35,456 galaxies). Given that our model from Section~\ref{sec:results} is trained using our full data set, we could expect its stellar mass predictions to be less accurate for this smaller subset of data, as it was constrained to model a larger data set. However we see that both the RMSE and the SMF of the new models are comparable to the one from our final model in section~\ref{sec:results}. The fact that our original model seems to do as well as these new ones suggest that our method is robust against sample size variations, and that it is able to deal effectively with overfitting.

Given that we see no improvement in accuracy using these new models, and given that they were not trained on our full data, we do not quote these new models as our final result, but keep the model of Section~\ref{sec:results} instead.

\section{Conclusions}
\label{conclusions}

In~\cite{icazalizaola2021sparse} we used a sparse regression methodology to fit the stellar mass of central galaxies as a function of properties of their host halo. In this paper we expand our study to cover a wider halo mass range, and to model the properties of satellite galaxies. The distinction between central and satellite galaxies relies on identifying subhalos as self-bound substructures within larger halos, for example by using the SUBFIND algorithm. This classification is uncertain and may be inconsistent for the same subhalos in adjacent snapshots outputs. We therefore explored whether we need to make a fundamental distinction between halos and subhalos. With this in mind, we use the maximum mass that a halo has ever reached during its evolution, denoted $\mathbf{Max}(M_\text{{total}}(z))$ and use this in place of the final (sub)halo mass at $z=0$. Given that central galaxies grow monotonically then $\mathbf{Max}(M_\text{{total}}(z))\sim M(z=0)$ and this results in little change. In subhalos, however,  it correspond to the mass of their main progenitor before merging with their central halo. In order to quantify the prior growth history of the halo, we define a set of formation criteria parameters, that measure the redshift at which a halo has formed a given percentage of its maximal mass and before it reaches $\mathbf{Max}(M_\text{{total}}(z))$.

Our data is taken from the EAGLE hydrodynamical simulation. In order to avoid selection biases when predicting stellar mass, we use a bijective matching between the EAGLE hydrodynamical simulation and a DM only simulation with the same cosmology and initial conditions.
We select all galaxies that have a halo mass larger than $\mathbf{Max}(M_\text{{total}}(z)/{\rm M_\odot})>10^{10.66}$, this value corresponds to the threshold at which our matching methodology successfully matches more than 90 percent of all galaxies. We use a total of 35,456 galaxies, 9,967 of them live inside subhalos and 25,489 inside central halos. Because our sample has significantly increased the fraction of low-mass galaxies considered compared to our previous work \citep{icazalizaola2021sparse} , we weight residuals according to stellar mass, giving a larger incentive to the model to accurately fit less well represented galaxy masses.

We build our models only using information on the accretion history of the halo or subhalo and its maximum mass. Using these parameters our methodology seem to predict the stellar mass of galaxies in halos and subhalos with a singular model and without needing to distinguish between the two.  We note that there are other parameters that we have not tested for in our analysis that might break this symmetry, for example, the infall angle of subhalos, which is not defined for central halos, might improve our modelling of subhalos.

The SMF of our models agrees well with that of EAGLE at all stellar masses except at $\log_{10}(M^*/{\rm M_\odot})=10.5$ where our models tend to slightly under-predict the amount of galaxies when compared with the EAGLE simulation. This could be related to the stochasticity of baryonic processes that might alter the stellar mass of a galaxy, which could be hard to predict using parameters from a DM only simulation.  We also calculate the correlation functions of our models split by their predicted stellar mass, and find that they also agree well with the EAGLE correlation functions.
The model that combines central and satellite galaxies has comparable accuracy to the models in which central and satellites are treated independently, while using an overall smaller number of model parameters. This suggests that a binary classification is unnecessary and the stellar mass of both galaxy types can be predicted by suitable measurement of their halo mass history.

The SRM approach can be viewed as a machine learning algorithm. It can accurately model the stellar masses of EAGLE from the data itself and without requiring previous knowledge of physics behind the system. At the same time, the approach results in a prediction algorithm that is explicit and simple (compared with the solutions of other machine learning techniques), and the terms that are retained give physical insight into the important processes at work.

We have seen that the correlation function and the stellar mass function of our models agree well with the EAGLE data set. This is encouraging as both of these EAGLE statistics have been positively compared with observational data. For example, \cite{2015MNRAS.450.4486F} has shown that the EAGLE SMF at $z=0$ agrees reasonably well with the ones observed by the SDSS \citep{2009MNRAS.398.2177L} and GAMA \citep{2012MNRAS.421..621B} surveys. Similarly \cite{10.1093/mnras/stx1263} shows that the EAGLE correlation function reproduces observations accurately between $1 h^{-1} \text{Mpc}$ and $6 h^{-1} \text{Mpc}$. Additional statistics, like Counts-in-Cells and multipoles of the correlation function, were successfully reproduced by the models, but we leave to future work a more in depth discussion of their successes and limitations.

Our method compares favourably with the SHAM methodology from \cite{2016MNRAS.460.3100C}, with both models being able to reproduce well the correlation function of EAGLE at larger stellar masses with our SRM models being slightly more accurate on smaller scales.

We also compare our model with the one presented in \cite{lovell2021machine}, using ERT which is a highly accurate ML methodology. ERT makes accurate models but the resulting models are less explainable than our SRM models. Both methods reach comparable accuracy on the SMF predictions, with our model being slightly more accurate at smaller stellar masses. 
We find similar predictions for the projected correlation function of a stellar mass selected sample between both models. We note that \cite{lovell2021machine} data was trained using C-EAGLE zoom-in simulation data that is not identical to the EAGLE data used in training our model, which might explain some of the small differences seen in the accuracy of the predictions of both models.

Finally, we analyze the inclusion of additional halo properties into our methodology. This is done by building new models with some of the halo parameters used in other successful ML models. We run five new models to account for differences due to variance in our methodology, which increases due to the correlations between the new parameters. We find no improvement in accuracy which suggests that any information provided by the new parameters was already present in our standard parametrization. We find a slight reduction in the number of surviving coefficients, which suggests that some parameters, like $\rm lgV_{peak}$ and $R_{1/2}$, are possibly more efficient at summarising some of the relevant information required to described the SMHM relation. However, the number of free parameters varies between five and nine depending on the model realisation, which complicates significantly the model interpretation, one of the underlying aims of this SRM methodology. Due to this fact along with the reduced stability of the model as evidenced by the increase in scatter on the predicted SMFs (Fig. ~\ref{Diff_param_smf}), we do not quote these new models as final result.

All of this suggests that our methodology could be a promising approach to populate N-body simulations with galaxies of the correct stellar mass and spatial distribution. 

However several complications will make this an interesting challenge. First, EAGLE is run in a comparatively small volume with respect to other DM simulations which means that the number of massive halos is comparatively small and it will be necessary to test the accuracy of the resulting SMF at the larger stellar masses. Second, larger simulations normally produce large amounts of output data, which generates challenges in storing the necessary halo history to build merger trees, some simulations either save only a small number of redshifts or no halo evolution information at all. Finally, the distribution of our required input halo parameters such as $\rm lgM_{ max}$ or $FC_i$ might differ from simulation to simulation. All of these reasons make populating larger simulations with galaxies using our methodology a challenging endeavour that we will explore in more depth in future papers.

Our ultimate goal is to generate mock catalogues that provide an accurate representation of the observed universe. An attractive idea is to iterate on the coefficients of the terms selected by comparison to EAGLE (or another hydrodynamic simulation), creating an even closer match to target observations. This would retain the same physical processes, but accept that their relative importance might differ between the true Universe and the simulation used for the training.

\section*{Acknowledgements}
We thank the anonymous referee for their very insightful comments that improved the paper. We thank Arnau Quera-Bofarull for his help in making our code faster and more efficient. We thank Christopher Lovell and Jon\'as Chaves-Montero for sharing their data with us, enabling the detailed comparisons presented.
MIL is supported by a PhD Studentship from the Durham Centre for Doctoral Training in Data Intensive Science, funded by the UK Science and Technology Facilities Council (STFC, ST/P006744/1) and Durham University. MIL, RGB, PN and SMC acknowledge support from the Science and Technology Facilities Council (ST/P000541/1 and ST/T000244/1).
This work used the DiRAC@Durham facility managed by the Institute for Computational Cosmology on behalf of the STFC DiRAC HPC Facility (www.dirac.ac.uk). The equipment was funded by BEIS capital funding via STFC capital grants ST/K00042X/1, ST/P002293/1 and ST/R002371/1, Durham University and STFC operations grant ST/R000832/1. DiRAC is part of the National e-Infrastructure. Some of the numerical calculations in this work were done in the high performance computer cluster of the Korea Astronomy and Space  Science Institute.

\section{Data Availability}

The data used in this work can be shared if requested from the authors. The data from the EAGLE simulations has been publicly released, see  \cite{McAlpine_2016}.




\bibliographystyle{mnras}
\bibliography{SMHM} 




\appendix

\section{Matching failures}
\label{Matching_failures}

As mentioned in Section~\ref{subsec:matching}, the matching success rate of satellite galaxies is around $80\%$. With this in mind, we decided to apply our model to all halos in the DM only simulation (match and unmatched) and compare the resulting statistics to the ones obtained from all galaxies in the EAGLE hydrodynamical simulation.

\begin{figure}
\includegraphics[width=85mm,height=85mm]{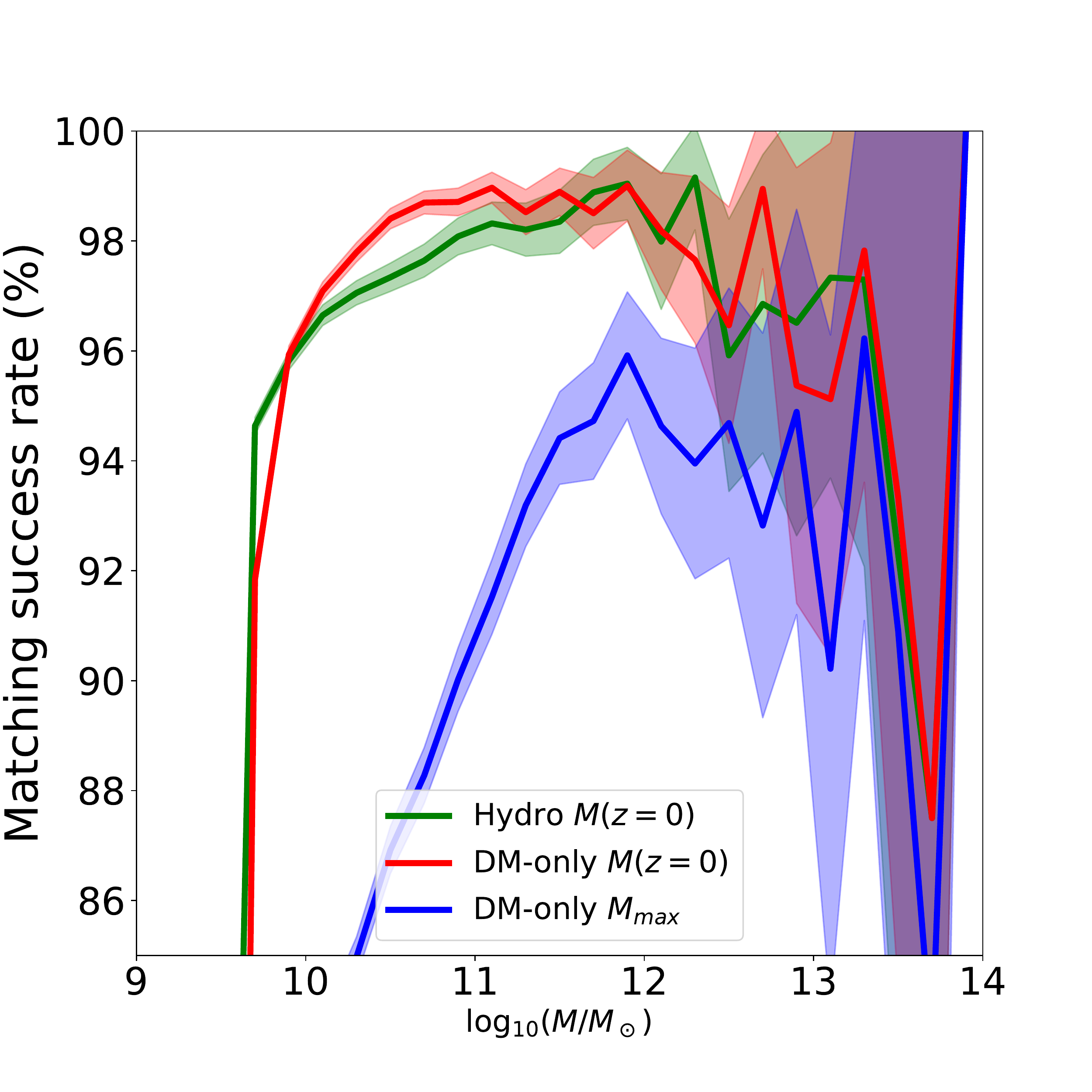}
\caption{The success rate of the matching methodology for halos in the hydrodynamical simulation (green line) and in the DM only simulation (red and blue lines) as a function of halo mass. For the green and red lines, the halo mass is $M_{\rm  total}$ at $z=0$, while for the blue line it is $M_{\rm max}$, the mass parameter used by our SRM model. The coloured shadings show the error on the matching rate assuming binomial statistics.}

\label{Match_succes}
\end{figure}

Fig.~\ref{Match_succes} shows the success rate of the matching algorithm as a function of the halo mass $M_{\rm  total}$ at $z=0$ for both the EAGLE-DMO simulation (red line) and the EAGLE hydrodynamical simulation (green line). For halos larger than $\log_{10}(M_{\rm total}/{\rm M_\odot})=11$ the percentage of unmatched halos is small ($< 2\%$) and similar across both simulations. This suggests that most of the halos that were unmatched in the hydrodynamical simulation did have an equivalent halo in the DM only simulation, but the algorithm had trouble matching them. This justifies the decision made in section~\ref{subsec:matching} to compare models applied to all halos (match and unmatched) in the DM only simulation tox all halos (match and unmatched) in the hydrodynamical simulation.

The matching algorithm runs at $z=0$ and therefore the lines at this redshift are adequate to show the success rate of the algorithm. However, we select our halo sample using $M_{\rm max}$, which is the maximum mass reached by the halo at any redshift. The matching success rate as a function of $M_{\rm max}$ is shown as the blue line in Fig.~\ref{Match_succes}. The matching success rate as function of $M_{\rm max}$ is smaller than when considering $M_{\rm  total}$ at $z=0$. The success rate is around $88\%$ at $\log_{10}(M_{\rm max}/{\rm M_\odot})=10.66$ (our mas cut), and grows to around  $94\%$ at $\log_{10}(M_{\rm max}/{\rm M_\odot})=11.5$. These differences in success rate are due to a significant fraction of halos being disrupted after a merger. These disrupted halos would be smaller at $z=0$ than at the redshift of their maximum mass, and therefore their probability of being matched decreases. This is consistent with other works that have found that an accurate measurement of $M_{\rm max}$ requires higher numerical resolution.

\section{Comparing with $V_{\rm peak}$}
\label{Vpeak}

In this section, we explore which one of the two halo properties $V_{\rm peak}$ and $M_{\rm max}$ would be a better input for our models. As stated in Section~\ref{subsec:mass_methodology}, the consensus is that stellar mass models that use properties correlated with the circular velocity profile of halos, like $V_\text{max}$ and $V_{\rm peak}$, tend to outperform those based on the mass of the halo. This is due to $V_{max}$ being a good representation of the inner part of the halo, which affects galaxies more directly and is less sensitive to mass striping. However, we note that the evolutionary history of the halo is well tracked in our SRM model due to our definition of $M_{\rm max}$ and the inclusion of formation criteria parameters. Therefore it is not trivial to know which of the two properties will perform better in our model.

We run our combined model from Table~\ref{Results_table}, but substituting ${\rm lgM_{ max}}$ for the unitless parameter ${\rm lgV_{\rm peak}}$ defined in Eq.~\ref{Vpeak_definition}.

\begin{figure}
\includegraphics[width=85mm,height=85mm]{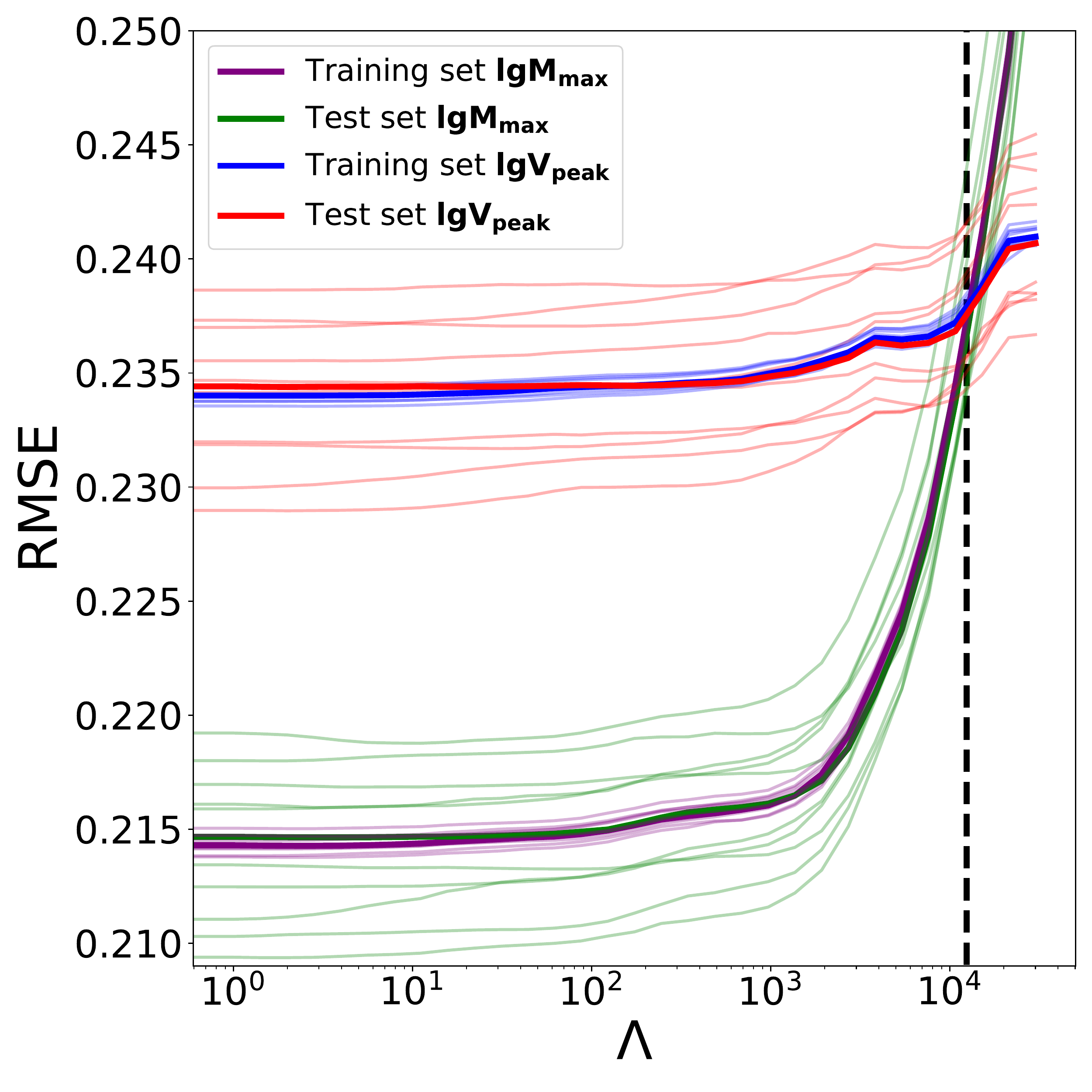}
\caption{RMSE reached by our algorithm at different values of the hyperparameter $\Lambda$ (Eq.~\ref{LASSO}) for the test and training sets of the k-fold method. The RMSE from the combined model of Table~\ref{Results_table} with the free parameter ${\rm lgM_{max}}$ are shown with green and purple lines, while those from a new model with ${\rm lgV_{\rm peak}}$ as a free parameter are shown with blue and red lines. The thin lines represent the RMSE for each of the k=10 individual data sets of the k-fold method and the thick lines show their mean value. The vertical black dashed line correspond to the $\Lambda$ value for which the model accuracy, described by the RMSE, is the same for both set of models.}
\label{V_max_plot}
\end{figure}

As mentioned in Section~\ref{methodology}  the optimal value of the hyperparameter $\Lambda$ from Eq.~\ref{LASSO} is found using a k-fold method, where the data is separated into a training set and a test set k-times. We examine how well a model fitted to the training sets at different values of $\Lambda$ predicts the test sets. We refer the reader to \cite{icazalizaola2021sparse} for an in-depth discussion of this process. Fig.~\ref{V_max_plot} shows the RMSE resulting from the exploration of the $\Lambda$ space for both models and the training and test sets of all k-folds. The figure shows how models that use ${\rm lgM_{ max}}$ as a parameter are more accurate than those using ${\rm lgV_{\rm peak}}$ for both training and test sets. We note that in this comparison the same set of formation criteria parameters were considered by both set of models and it is within this specific modelling context that we draw our conclusions. 
 
The models that use ${\rm lgV_{\rm peak}}$ are less accurate, but they are simpler than the one with ${\rm lgM_{ max}}$, as the former require only six parameters. We can build a simpler ${\rm lgM_{ max}}$ model by increasing the magnitude of $\Lambda$ beyond its nominal optimal value. The black dashed line in Fig.~\ref{V_max_plot} shows the value of $\Lambda$ at which a model built with ${\rm lgM_{ max}}$ reaches the same accuracy as the one with ${\rm lgV_{\rm peak}}$. The resulting ${\rm lgM_{ max}}$ model built with this $\Lambda$ contains seven free parameters, which is very comparable with the six of the ${\rm lgV_{\rm peak}}$ model. With this in mind, we conclude that models built with ${\rm lgM_{ max}}$ are more accurate and can be as simple as models built with ${\rm lgV_{\rm peak}}$. This justifies our selection of ${\rm lgM_{ max}}$ as the mass parameter used in this work.

\section{Correlated Parameters}
\label{Correlated Parameters}

\begin{figure*}
\includegraphics[width=\textwidth,trim={6cm 6cm 6cm 6cm},clip]{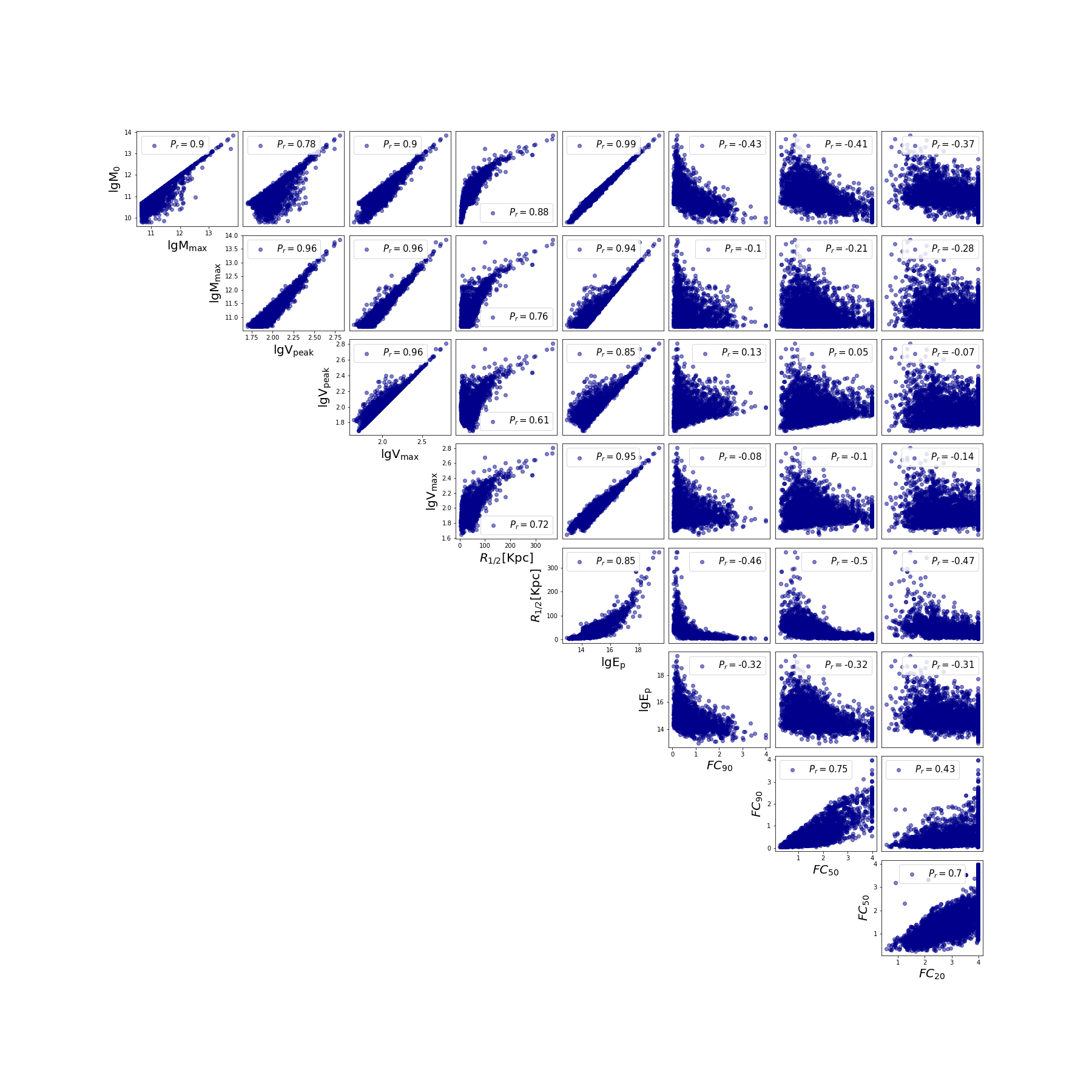}
\caption{Correlations of most halo properties used in this work, as indicated by the axis labels of each panel. The Pearson correlation coefficient is indicated in each panel, as a measure of how correlated two halo properties are.}
\label{correlation_plot}
\end{figure*}

Figure~\ref{correlation_plot} shows the correlations of most halo properties used throughout this work, built from the 4,000 halos considered in Section~\ref{OtherParameters}. For clarity, the parameters $FC_{30}$ and $FC_{70}$ have been omitted, as they show similar correlation trends to the other three formation criteria parameters already included. Each panel includes $P_r$, the value of the Pearson correlation coefficient\footnote{The Pearson correlation coefficient is defined as the ratio of the covariance of the parameters with the product of their standard deviations.} for each pair of halo properties. The closer the absolute value of this coefficient is to unity, the more linearly correlated those two parameters are.

Figure~\ref{correlation_plot} shows that parameters can be divided into two subgroups of correlated halo properties:
\begin{itemize}
    \item The first subgroup includes $\rm lgM_0$, $\rm lgM_{ max}$, $\rm lgV_{peak}$, $\rm lgV_{max}$, $R_{1/2}$, and $\rm lgE_p$. They are all strongly correlated with each other, with |$P_r$| around than 0.9 typically. 
    \item The formation criteria parameters $FC_i$ form the second subgroup. Their correlations, as measured by the Pearson coefficient, are weaker than those within the first group, with |$P_r$| less than 0.7 typically.
    
\end{itemize}

Out of all of the parameters in the first group, $R_{1/2}$ is the least correlated with the rest, with Pearson coefficients between 0.61 and 0.88 with respect to the rest of the halo properties of this subgroup. This might explains why most of the five models of Section~\ref{OtherParameters} select a small but noticeable contribution from $R_{1/2}$ after already having strong contributions from $\rm lgM_{max}$ and $\rm lgV_{peak}$. It is also noticeable how correlated $\rm lgM_0$ and $\rm lgE_p$ are with each other, with a correlation coefficient of 0.99. This suggests that the information that they could provide to a model is almost identical.
The strong correlation between both parameters comes from the similarities in the way they are defined and computed in EAGLE \citep[see Appendix~D in][]{McAlpine_2016}.

The fact that the parameters of this first subgroup are so correlated with each other might explain why the model of Section~\ref{results_main}, which is built using only $\rm lgM_{max}$ has comparable accuracy to the models of Section~\ref{OtherParameters}, which are made using all of the six parameters of this subgroup.
    
The weaker correlations observed between parameters in the second group, i.e.\ the formation criteria parameters $FC_i$, suggest that the information held by those halo properties is somewhat more unique, especially when compared to the halo properties of the first subgroup. This might explain why all of the models presented in this work select several parameters of this subgroup simultaneously.
 
As we discuss in detail in Section~\ref{OtherParameters}, the inclusion of correlated parameters adds stochastisity to our resulting models. This can be seen in models selecting very different collections of surviving coefficients when built with different subsets of training data. As mentioned this is due to correlated parameters making the parameter space non-convex, with several local minima. Dealing with correlated parameters is something that would need to be implemented into our methodology in future work, if uniqueness of the solution and maximal parameter reduction is a priority.

\bsp	
\label{lastpage}
\end{document}